\documentclass[preprint,authoryear,12pt]{elsarticle}

\usepackage{epsfig}

\usepackage{amssymb}

\usepackage[ps2pdf,%
a4paper=true,%
breaklinks=true,%
colorlinks=true,%
pdfauthor={First Author et al.},%
pdftitle={Template for manuscripts in Advances in Space Research}%
]{hyperref}

\journal{Advances in Space Research}

\begin{document}

\begin{frontmatter}



\title{Comptonization efficiencies of the variability classes of GRS 1915+105}


\author{Partha Sarathi Pal\corref{cor}\fnref{footnote1}}
\address{S. N. Bose National Centre For Basic Sciences, Kolkata.}
\cortext[cor]{Corresponding author}
\fntext[footnote1]{S. N. Bose National Centre For Basic Sciences, Kolkata.
JD Block, Salt Lake, Kolkata - 700098, India.
Tel: +91-33-2335-5706/7/8 Extn: 374,
Fax: +91-33-2335-9176.
}
\ead{parthasarathi.pal@gmail.com, partha.sarathi@boson.bose.res.in}


\author{Sandip K. Chakrabarti\fnref{footnote1}}
\ead{chakraba@bose.res.in}

\author{Anuj Nandi\fnref{footnote3}}
\address{ISRO Satellite Centre, Bangalore}
\fntext[footnote3]{Space Astronomy Group, INSAT Building,
ISRO Satellite Centre, Bangalore - 560017, India}
\ead{anuj@isac.gov.in}

\begin{abstract}
The Galactic microquasar GRS 1915+105 exhibits at least seventeen types of variability classes. 
Intra and inter class transitions are reported to be observed within seconds to hours.
Since the observation was not continuous, these classes 
appeared to be exhibited in a random order. Our goal is to predict a sequence of these classes.
In this paper, we compute the ratio of the photon counts obtained from the 
power-law component and the blackbody component of each class 
and call this ratio as the `Comptonizing efficiency' (CE) of that class.
We sequence the classes in the ascending order of CE and find that this sequence
matches with a few class transitions observed by RXTE satellite and IXAE instruments on board IRS-P3.
A change in CE corresponds to a change in the optical depth of the Compton cloud.
Our result implies that the optical depth of the Compton cloud gradually rises as the variability class becomes harder.

\end{abstract}

\begin{keyword}
{Black Holes, Accretion disk, X-rays, Radiation mechanism}
\end{keyword}

\end{frontmatter}

\parindent=0.5 cm

\section{Introduction}

The enigmatic stellar mass black hole binary GRS 1915+105 \citep{harla} 
was first discovered in 1992 by the WATCH detectors \citep{castro92} as a transient source 
with a significant variability in X-rays \citep{castro94}. In the RXTE era, 
GRS 1915+105 was monitored over thousands of times in X-rays and 
the data reveal a very unique nature of this compact object, in that, the nature of the
light curve was highly unpredictable. The radio observation with VLA suggests apparent
superluminal nature of its radio jets. Radio observation constrains that its maximum distance is 
no more than $13.5$ kpc and that the jet axis makes an angle of $70^{\circ}$ 
with the line of sight \citep{mira94}. 

Continuous X-ray observation of GRS 1915+105 reveals that the X-ray intensity of the source changes 
peculiarly in a variety of timescales ranging from seconds to days \citep{greiner96, morgan97}. 
Small scale variabilities of GRS 1915+105 are identified with local variation of the inner disk 
\citep{nan00, skc00a, migliari03}. Several observers have reported that this object exhibited many 
types of variability classes \citep{yad99, ra00b, bel, skc00b, nai02a}. 
Depending on the variation of photons in different arbitrary energy bands (hardness ratio) 
and conventional color-color diagram of GRS 1915+105, the X-ray variability of the source was found to have
fifteen ($\alpha,\ \beta,\ \gamma,\ \delta,\ \phi,\ \chi_1, \chi_2, \chi_3, \chi_4,\ \mu,\
\nu,\ \lambda,\ \kappa,\ \rho,\ \theta$) classes. In a 1999 observation of RXTE, 
the existence of a new class $\omega$ was reported \citep{wolt02, nai02a}. 
In 2003, one other class, namely, $\xi$ was reported \citep{hann}.
In the so-called $\chi$ (i.e., $\chi_1$ to $\chi_4$) 
class, the strong variability as is found in other classes is absent. 
The classes named $\chi_1, \chi_3$, $\beta$ and $\theta$ are  found to be 
associated with strong radio jets \citep{nai00, vad01, vad03}. 
\citet{ueda09} presented the results from simultaneous Chandra HETGS and RXTE 
observations of GRS 1915+105 in its quasi-stable ``soft state'' (the so-called State A). 
The continuum spectra obtained with RXTE in the $3-25$ keV band 
is found to be fitted assuming a thermal Comptonization with an electron temperature of 
$\sim 4$ keV and an optical depth of $\sim 5$ from the seed photons from the standard disk extending 
down to $(4-7) \ r_g$. Most of the radiation energy appears to be
produced in the hot electron cloud which completely covers the inner part of the disk. 
\citet{ne11} studied in detail the $\rho$ variability class
whose spectral and timing properties were found to be consistent with the radiation pressure instability.
At small scales $(1-10)\ r_g$, they detect a burst of bremsstrahlung 
emission that appears to occur when a portion of the inner accretion disk evaporates due to 
radiation pressure. Jet activity, as inferred from the appearance of a short X-ray hard state, 
seems to be happening at times near minimum luminosity, with a duty cycle of $\sim 10$ \%. 
\citet{skc00b} suggested that not only in $\rho$, in all the classes, some outflows should be
present when the state is harder \citep{ne09}, since only in harder states
the Compton cloud, which is also the base of the jet in this model, is significantly hot.
\citet{naya00} investigated different accretion disk models and viscosity 
prescriptions and concluded that the X-ray observations clearly 
require a quasi-stable accretion disk solution at a high accretion rate 
at which the radiation pressure begins to dominate. This excludes the standard 
$\alpha$ viscosity prescription. They devise a simplified 
model of a disk with a corona and a modified viscosity law that has a quasi-stable upper branch. Their
model appears to account for several gross observational features of GRS 1915+105, 
including its overall cyclic behavior on timescales of $\sim 100-1000$s. The inclusion of a jet 
allowed them to reproduce several additional observed features as well. \citet{cab09}  
reported that in response to major changes in the mass accretion rate within the inner accretion flow, 
the black hole binary transients undergo dramatic evolution in their X-ray timing and 
spectral behavior during outbursts.

Quasi-Periodic Oscillations (QPOs) are observed in GRS 1915+105 in a wide range of frequencies. 
QPOs in this source are associated with different types of X-ray variabilities and their timing properties 
are correlated with spectral features \citep{mun99, sob99, rod02, vig03, dun10}. 
The origin of QPO frequencies between $0.5$ to $10$ Hz is identified to be due to 
the oscillation of the Compton cloud, presumably the post-shock 
region of the low angular momentum (sub-Keplerian) flow \citep{skc00a, ra00a}. 
There are several other models
of QPOs in the literature. For instance, \citet{mik09} state that the low-frequency 
quasi-periodic X-ray oscillations observed microquasars are correlated to, but do not originate at, 
the physical radius of the inner edge of the accretion disk. By analyzing several
data sets including $\theta$, $\beta$, and $\alpha$ they find that accretion-ejection
instability (AEI) model may explain the nature of correlations between the frequency and radius.

While a large number of reports appear in the literature on GRS 1915+105,
to our knowledge, there is no work which actually asked the question: are these variability
classes arbitrary, or they appear in a given sequence? The problem lies in the fact that no satellite 
continuously observed GRS 1915+105. Sporadic observations caught the object in sporadic
classes, and thus no specific sequence of the variability classes has been identified. 
In the present paper, we compute a simple parameter obtainable directly from observations.
This parameter, which we term as the Comptonization efficiency or CE, is simply
the ratio between the number of photons in the power-law component  
and number of photons in the black body component. This is not simply `a' generic hardness ratio
that is often reported in the literature, since the energy bands of the soft and hard photons
in our consideration vary dynamically from class to class as decided by 
the best $\chi^2$ fit. We find that, when averaged over a full cycle 
(whenever available) or over a sufficiently long time period (whenever there 
is no obvious periodicity in the light curve), every class is characterized 
by more or less a unique CE. When these CEs are placed in an ascending order, 
a sequence of the classes is generated. It appears that the direct class transitions which 
have been observed so far, do generally take place between the nearest neighboring classes of this
particular sequence that we propose. Of course, if the formation of a class 
(such as the $\theta$, $\beta$ and $\alpha$) requires special physical condition, 
such as the presence of a strong magnetic field, 
that class may be skipped and the transition could be to the next nearest neighbor. 
Since the number of photons in the power-law component depends on the degree of interception by the      
so-called hot electron cloud or Compton cloud \citep{sun80, sun85}, CE thus parameterizes different 
classes by the average optical depth of the Compton cloud. To accomplish the 
computation of running value of  CE, we separate out the photons $\gamma_{BB}$ of the black body component 
and the photons $\gamma_{PL}$ from the power-law component and take the running ratio CE 
as a function of time to study how the Compton cloud itself varies in a short time 
scale. Our finding reveals that the Compton cloud is highly dynamic. We present possible 
scenarios of what might be happening inside it in different variability classes.  
In this paper, along with the dynamical evolution of CE, we also compute the spectrum 
and the power density spectrum (PDS) of each of the classes. 
A preliminary report is presented in \citet{pal08}. 

The paper is organized as follows: in the next Section, we present a general discussion of 
the observation, our criteria of selection of data and the analysis technique. 
In \S 3, we discuss the procedure of calculation that is adopted to calculate the photon numbers.
In \S 4, the results are presented. In \S 5, we show that the Comptonizing efficiency may be 
a key factor to distinguish among various classes. Finally, in \S 6, we make concluding remarks.

\section{Observation \& Data Analysis}

The RXTE science data is taken from the NASA HEASARC data archive for analysis. We have chosen
the data procured in 1996-97 by RXTE as in this period GRS 1915+105 has shown almost
all types of variabilities in X-rays. Subsequently, in 1999, another class, namely $\omega$ was
seen. In the present paper, since we are interested in sequencing these seventeen classes,
we will rename them as follows: I=$\phi$; II=$\delta$; III=$\gamma$; IV=$\omega$; V=$\mu$;
VI=$\nu$; VII=$\lambda$; VIII=$\kappa$; IX=$\rho$; X=$\xi$; XI=$\beta$; XII=$\alpha$; XIII=$\theta$; 
XIV=$\chi_2$; XV=$\chi_4$; XVI=$\chi_1$; XVII=$\chi_3$. This renaming was done on purpose,
since it will turn out that this is the sequence in which average CE ($<CE>$) increases.

During the data analysis we exclude the data collected for elevation angles less than $10^{\circ}$, for
offset greater than $0.02^{\circ}$ and those acquired during the South Atlantic Anomaly (SAA) passage. 
In Table 1, the details of the data selection and ObsIDs are given which we analyzed in this paper. 
In the first column, we give the observational ID, in the second column, we present the class number
(new and old), and finally, in the third column, we present the date of observation.  

\begin{table}
\centering
\begin{tabular}{ccc}
Obs-Id & Class & Date \\
\hline\hline
10408-01-19-00$^*$ & I ($\phi$) &29-06-1996\\
10408-01-09-00 & I ($\phi$) &29-05-1996\\
10408-01-18-00$^*$ &II ($\delta$) &25-06-1996\\
20402-01-41-00 & II ($\delta$) &19-08-1997\\
20402-01-37-00$^*$ & III ($\gamma$) &17-07-1997\\
20402-01-56-00 & III ($\gamma$) &22-11-1997\\
40703-01-27-00$^*$ & IV ($\omega$) &23-08-1999\\
40403-01-07-00 & IV ($\omega$) &23-04-1999\\
10408-01-36-00$^*$ & V ($\mu$) &28-09-1996\\
20402-01-53-01 & V ($\mu$) &05-11-1997\\
10408-01-40-00$^*$ & VI ($\nu$) &13-10-1996\\
20402-01-02-02 & VI ($\nu$) &14-11-1996\\
20402-01-36-00$^*$ & VII ($\lambda$) &10-07-1997\\
10408-01-38-00 & VII ($\lambda$) &07-10-1996\\
20402-01-33-00$^*$ & VIII ($\kappa$) &18-06-1997\\
20402-01-35-00 & VIII ($\kappa$) &07-07-1997\\
20402-01-03-00$^*$ & IX ($\rho$) &03-06-1997\\
20402-01-31-00 & IX ($\rho$) &19-11-1996\\
80127-01-02-00$^*$ & X ($\xi$) & 2003-03-07\\
10408-01-10-00$^*$ & XI ($\beta$) &26-05-1996\\
20402-01-44-00 & XI($\beta$)  &31-08-1997\\
20187-02-01-00$^*$ & XII ($\alpha$)&07-05-1997\\
20402-01-30-01 & XII ($\alpha$) &28-05-1997\\
10408-01-15-00$^*$ & XIII ($\theta$) &16-06-1996\\
20402-01-45-02 & XIII ($\theta$) &05-09-1997\\
20402-01-16-00$^*$ & XIV ($\chi_2$)&22-02-1997\\
20402-01-05-00 & XIV ($\chi_2$) &04-12-1996\\
20402-01-25-00$^*$ & XV ($\chi_4$) &19-04-1997\\
10408-01-33-00 & XV ($\chi_4$) &07-09-1996\\
10408-01-23-00$^*$ & XVI ($\chi_1$) &14-07-1996\\
10408-01-30-00 & XVI ($\chi_1$) &18-08-1996\\
20402-01-50-00$^*$ & XVII ($\chi_3$) &14-10-1997\\
20402-01-51-00 & XVII ($\chi_3$) &22-10-1997\\
\end{tabular}
\caption{The ObsIDs and the dates of the RXTE data analyzed in this paper. $^*$ represents the result shown 
in Fig. 3 of this paper.}
\label{table:1}      
\end{table}
 
Here we discuss how we analyzed the temporal and spectral properties of the data.

\subsection{Timing Analysis}
In the timing analysis of the RXTE/PCA data we use ``binned mode'' data which was available
for 0-35 channels only, with a time resolution of $2^{-7}$ sec. We use the `event mode' data with a time
resolution of $2^{-8}$ sec for the rest of the channels. We restrict ourselves in the energy
range of $\sim 2-40$ keV for the timing analysis of PCA data. After extracting the light
curves using standard tasks from two different modes, we add them by the FTOOLS
task ``lcmath'' to have the whole energy range light curve ($2$ to $40$ keV). The Power Density
Spectrum (PDS) is generated by the standard FTOOLS task ``powspec'' with a suitable normalization.
The data is re-binned to $0.01$s time resolution to obtain a Nyquist frequency of $50$ Hz as the power
beyond this is found to be insignificant. PDSs are normalized to give the squared {\it rms}
fractional variability per Hertz.

To have the timing evolution of QPO features in different classes, we generate PCA
light curves of $2$ to $40$keV photons with $0.01$s time bin. We use the science data
accordingly from the `binned mode' and the `event mode' for $10$s time interval in each step.
This light curve is then used to make PDS using the ¨powspec¨ task. The dynamic PDS is made
by considering the shift of time interval by $1$s. The selection of time interval is done by
FTOOLS task `timetrans'. The PDSs are plotted accordingly to see the variation of QPO features
in a particular class of a few hundred seconds of observation.

\subsection{Spectral Analysis}
	
Spectral analysis of the PCA data is done by using ``standard2'' mode data which
have $16$ sec time resolution and we constrained our energy selection up to $40$ keV to match with
the timing analysis. The source spectrum is generated using FTOOLS task ``SAEXTRCT'' with $16$ sec
time bin from ``standard2'' data.  The background fits file is generated from the ``standard2'' fits 
file by the FTOOLS task ``runpcabackest'' with the standard FILTER file provided with the package.
The background source spectrum is generated using FTOOLS task ``SAEXTRCT'' with $16$ sec time bin from
background fits file. The standard FTOOLS task ``pcarsp'' is used to generate 
the response file with appropriate detector information.
The spectral analysis and modeling was performed using XSPEC (v.12) astrophysical fitting
package. For the model fitting of the PCA spectra, we have used a systematic error of $0.5\%$. The
spectra are fitted with diskbb and power-law model along with $6.0\times 10^{22} cm^{-2}$ hydrogen
column absorption \citep{mun99} and we used the Gaussian for iron line as required for the best fitting.
\citet{morri83} reported that mainly affects the photons from $0.03$ to $10.0$ keV. It is important to consider
hydrogen column absorption for RXTE analysis because $2.0-10.0$ keV photons contain several informations.
As we are interested about the changes in the frame of the compact object we have chosen the $n_H$ carefully. 
While choosing the model to extract the number of photons, we need to be careful. First of all, one single
model must be able to fit spectra of all the seventeen classes with sufficient accuracy, otherwise, the comparison
of the Comptonization efficiency becomes meaningless. Second: the model must have the 
flexibility to provide soft-photons in a dynamically obtained energy band,
the limit of which is to be determined automatically so as to get the best $\chi^2$. While there are a 
number of models in the literature such as ComptST, SIMPL etc., we adopt the method used by \citet{sob99}
to obtain the spectral parameters as it satisfies both the criteria mentioned above. 
We also point out that at this stage, we do not consider the effects of hardening factor 
\citep{shi95} arising out of Comptonization by an optically thin haze over the accretion disk.
In Figs. 1(a-b), we present an example (Obs. ID 20402-01-16-00) of 
why \citet{sob99} is useful in our context.
Both the Figures represent the analysis of the same spectrum of a data in XIV class.
In Fig. 1(a), the spectrum is fitted by the standard method with the {\it diskbb} and power-law model.
From this unfolded spectrum, it is observed that the {\it diskbb} component is weaker than the power-law
component. However, if we try to fit this spectrum only with a power-law component then similar fitting 
parameters are also obtained and the $\chi^2$ value remained similar. So the ratio of the power-law photon number
to diskbb photon number would have no meaning. On the contrary, in Fig. 1(b), we fit 
the spectrum with the \citet{sob99} method. This gives a similar $\chi^2$ and the unfolded spectrum is shown.  
The ratio CE is calculable without any confusion. The results of the analysis of all 
the classes using this method are given in Table 2 \& 3 which will be described later.
We have verified that for the intermediate classes, the parameters obtained from both 
the fitting methods remain similar. However, since in the present paper, 
we are dealing with all the variability classes with a significant spectral difference,  
we adopted the method of \citet{sob99} to analyze all the classes using a 
single procedure. Another advantage of this method is that we can
compute the upper limit of energy $dbb_e$ to be used for the {\it diskbb} model dynamically 
in every time bin of the spectrum and thus obtained the dynamical 
hardness ratio, i.e., the Comptonization efficiency, CE. 
We have obtained $dbb_e$ by fitting the spectrum with {\it diskbb} only
and changing the upper energy limit $E_{fit}$ till we have a 
$\tilde{\chi}^2$ value $\approx 1$. This yields $dbb_e=E_{fit}$. 
From Tables 2 \& 3, we see the variation of $dbb_e$ for all the classes.
In Table 2, (which is for {\it diskbb} plus power-law model) 
the column 1 gives the variability class with 'h' and 's' 
representing the harder (burst-off) and softer (burst-on) states respectively.
Columns 2 and 3 give the black body temperature 
and the upper energy limit $dbb_e$ obtained from the fit. 
Column 4 gives the $\chi^2(dof)$ for the diskbb spectral model fitting. Column 5 gives the 
number of blackbody photons between $dbb_e$ and $0.1$keV. Column 6 gives the 
power-law index. Column 7 contains the number of Comptonized photons 
in the range of $3 \times T_{in}$ to $40$keV. Last two columns give the 
Comptonizing efficiency CE and the $\chi^2(dof)$ higher energy range. Table 3 is for {\it diskbb}
plus {\it ComptST} model. Here, most of the columns have the same meaning as in Table 2. However, instead
of the power-law index $\alpha$, we presented the electron temperature $T_e$ and the optical depth $\tau$.
Note that it is not always possible to fit with the {\it COMPST} model. Hence some of the rows 
in Table 3 do not have entries. We have calculated error-bars at 90\% confidence level in each case.

To have the spectral evolution with time for each class, we have generated the PCA spectrum
($2.0$ to $40$keV) with a minimum of $16$s time interval along with the background spectrum and response
matrix. This procedure is repeated with every $16$s shift in time interval since the minimum
time resolution in `standard2' data is $16$s. We use `timetrans' task to select each step 
of time interval. Spectral evolution (dynamic representation) for a longer duration of observation 
($\sim 500$ to $2000$ sec) of each class is plotted to show the local spectral variation in each class. 

\section{Calculation Procedure}

In this Section, we discuss the process of calculation of photon numbers using the parameters obtained
from the spectral fit.

We fit each spectrum in XSPEC environment using the models and the constraints mentioned earlier.
The fitting parameters are used to calculate the black body photons from the standard disk and
the power-law photons from the Compton cloud. The number of black body photons
are obtained from the fitted parameters of the multi-color disk black body model \citep{mak86}. 
This is given by,
\begin{equation}
f(E)=\frac{8\pi}{3} r_{in}^2 \cos{i} \int_{T_{out}}^{T_{in}} (T/T_{in})^{-11/3}B(E,T)\,dT/T_{in},
\end{equation}
where, $B(E,T)=\frac{E^3}{(\exp{E/T}-1)}$ and $r_{in}$ can be calculated from,
\begin{equation}
K=(r_{in}/(D/10kpc))^2 \cos{i},
\end{equation}
where, $K$ is the normalization of the blackbody spectrum obtained after fitting, $r_{in}$ is the
inner radius of the accretion disk in $km$, $T_{in}$ is the temperature at $r_{in}$ in keV, $D$ is
the distance of the compact object in kpc and $i$ is the inclination angle of the accretion
disk. Here, both the energy and the temperature are in keV. The black body flux, $f(E)$ in
photons/s/keV is integrated between $0.1$ keV to the maximum energy $dbb_e$. This
gives us $\gamma_{BB}$, the rate of the number of the emitted black body photons.   

The Comptonized photons $\gamma_{PL}$ that are produced due to inverse-Comptonization of the soft 
black body photons by `hot' electrons in the Compton cloud are calculated by fitting with 
the power-law given below,
\begin{equation}
P(E)=N(E)^{-\alpha},
\end{equation}
where, $\alpha$ is the power-law index and $N$ is the total $photons/s/cm^2/keV$ at $1$keV.
It is reported in \citet{tit94}, that the Comptonization spectrum will have a peak at around
$3 \times T_{in}$. The power law equation is integrated from $3 \times T_{in}$ to $40$keV to
obtain the rate of emitted Comptonized photons. The Gaussian function $Ga(E)$ is incorporated
with a power-law equation for the presence of line emissions in the spectrum, 
whenever necessary is given below,
\begin{equation}
Ga(E)= L \left (\frac{1}{\sigma \sqrt{2\pi}} \right) \exp \left(-0.5 \left
(\frac {E-E_l}{\sigma}\right )^2 \right ),
\end{equation}
where, $E_l$ is the energy of emitted line in keV, $\sigma$ is the line width in keV and $L$ is the
total number of $photons/s/cm^2$ in the line. The variation of the Comptonizing Efficiency (CE) with
time is plotted to have an idea of how the Compton cloud may be changing its geometry.

Just for comparison, we tried to fit the data with another Comptonization model, namely, the {\it compST}
\citep{sun80}. The spectrum is given by,
\begin{eqnarray}
Cst(x)&=& \frac{\alpha(\alpha+3)}{2\alpha+3} \left( \frac{x}{x_0} \right)^{\alpha+3}, when~ 0\leq x \leq x_0, \nonumber\\
      &=& \frac{\alpha(\alpha+3){x_0}^\alpha}{\Gamma(2\alpha+4)}x^3 \exp(-x) I(x), when ~ x \geq x_0. \
\end{eqnarray}
Here, $I(x)= \int^{\infty}_0 t^{\alpha-1} \exp{-t} \left( 1+\frac{t}{x} \right)^{\alpha+3}$, 
$x=E/kT_e$, $x_0=kT_{in}/kT_e$ and $\alpha$ is the spectral index. This $\alpha$ is calculated
from $\alpha=\sqrt{\frac{9}{4}+\gamma}-\frac{3}{2}$, where $\gamma = \frac{\pi^2 m_e c^2}
{3(\tau+\frac{2}{3})^2 kT_e}$. $kT_e$ is the electron temperature and $\tau$ is the optical depth of the
electron cloud. In the computation, we exclude the hydrogen column absorption feature while calculating CE.
This is because we are interested in the photons which were emitted from the disk before
suffering any absorption. Thus the variation of CE along with other features (e.g., spectral
and QPO frequency variations) will reflect the {\it actual} radiative properties of the flow near a black hole.  

\begin{figure*}
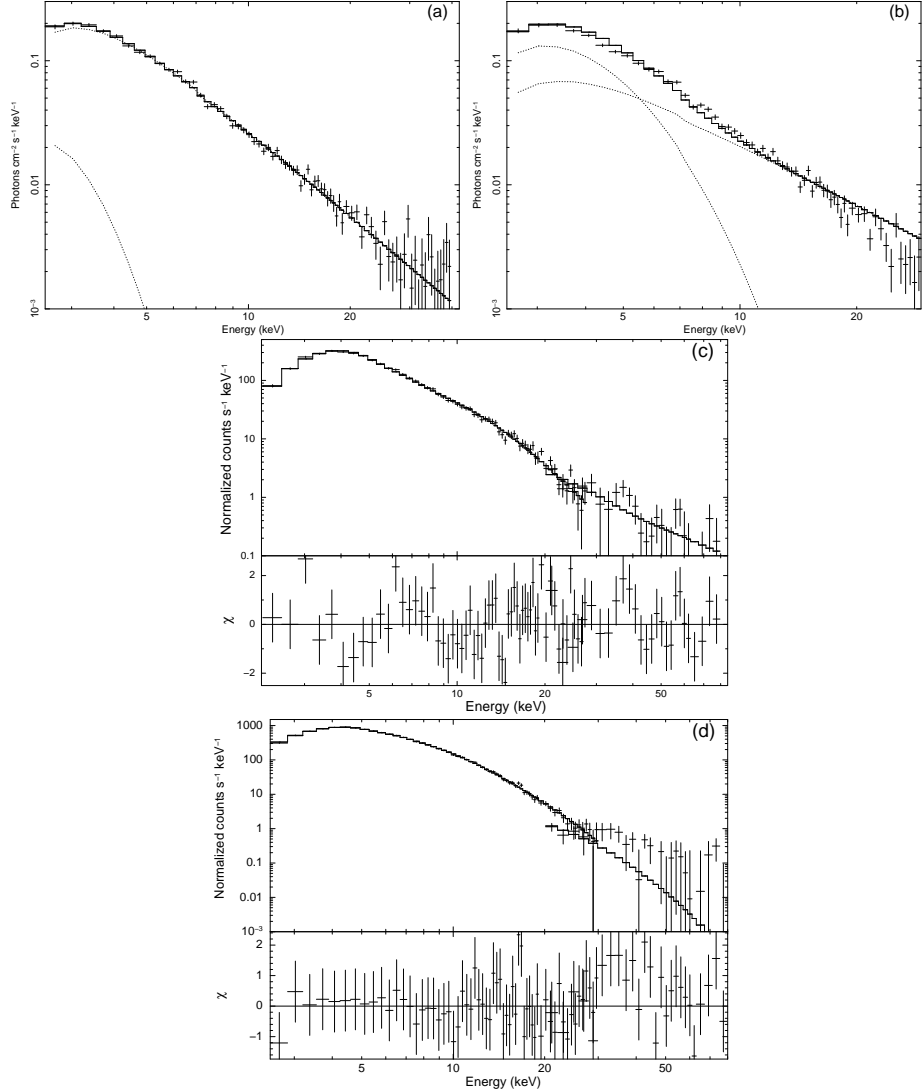

\centering
\includegraphics[angle=-90,width=6cm]{fig1a.ps}
\includegraphics[angle=-90,width=6cm]{fig1b.ps}
\includegraphics[angle=-90,width=7cm]{fig1c.ps}
\includegraphics[angle=-90,width=7cm]{fig1d.ps}
\caption{(a) The unfolded PCA spectrum of a data (Obs. ID 20402-01-16-00)
fitted with the {\it diskbb} and power-law with the standard method.  
(b) The unfolded PCA spectrum of (a) fitted with {\it diskbb} and power-law with the \citet{sob99} method.     
(c) The fitted spectrum of PCA \& HEXTE data (Obs. ID 20402-01-33-00) 
along with the {\it diskbb} and power-law components.
(d) The PCA \& HEXTE  sample spectrum of (c) fitted with {\it diskbb} and {\it compST} components.
\label{compare}}
\end{figure*}

In Figs. 1(c-d), we show the PCA and HEXTE spectrum of the data of the same Obs. ID (20402-01-33-00)
along with the fitted components. We also show the residuals to characterize the goodness of the fit.
In Fig. 1(c), the fitting is done with {\it diskbb} and power-law components. The photon
numbers and CE are calculated with the parameters obtained from the fitting.  
The black body photons are calculated with $T_{in}=1.15~^{+0.06}_{-0.05}$ keV. 
Here the $\tilde{\chi}^2$ is 1.1 with 10 degrees of freedoms. The calculated number of 
black body photons from $0.1-6.5$keV is $156.03~^{+17.10}_{-13.90}$ kphotons/s. In the same way, 
the power-law photons are calculated with the power-law index = $2.06~^{+0.15}_{-0.15}$.
The calculated number of the Comptonized photons between $3.45-40$ keV is $0.39~^{+0.06}_{-0.05}$ kphotons/s. 
The ratio between the power-law photon and the black body photon is $0.25~^{+0.08}_{-0.07}$\%. 
This means that only $0.25$\% of the soft photons are Comptonized by the Compton cloud.
In Fig. 1(c), we also include HEXTE data, and fitted with a power-law. But the spectral fit does not change 
in slope and the number of photons contributed by HEXTE region is so low that the value
of CE remains unchanged.
In Fig. 1(d), the same spectrum is fitted with {\it diskbb} and {\it compST} models.
The calculation with the {\it compST} model parameters
are found to be $kT_e = 6.41~^{+2.55}_{-1.25}$ and $\tau = 9.42~^{+2.88}_{-2.39}$. 
The number of Comptonized photons within the same range as in Fig. 1(c) is $0.41~^{+0.06}_{-0.05}$ kphotons/s. 
Thus the ratio between the {\it compST} photon and the {\it diskbb} photon is $0.26~^{+0.08}_{-0.07}$\%. 
In case of harder states the both PCA and HEXTE analysis give similar model parameters at the higher energies.
In case of softer states the higher energy region data has bigger error bars.  
This is true for several variability classes. We therefore ignore the HEXTE data in the rest of the paper to keep 
the analysis procedure similar for all the variability classes. 
It is to be noted that there is a scope of possible double counting of a very few photons in the region
where the blackbody and power-law components overlaps. We have verified that this does not amount to 
a change of more than $0.3$\% in CE.

\citet{ste11} presented an empirical model of Comptonization for fitting the spectra 
``SIMPL". This package has been implemented in XSPEC. We have carried out our fitting by using
this model also. The value of $\Gamma$ (photon power-law index) has been found to be
nearly equal (within errorbars) to the power-law index we obtained from using 
power-law model. The scattered fraction parameter FracSctr
of this model gives the amount of black body photons Comptonized by the electron cloud.
If we calculate CE within a specific energy range (say $2-40$keV) 
we get a similar value as FracSctr parameter. However, we are not only interested in the $2-40$keV range,
but also in the range in which the data is not obtained by RXTE. That is to say, we like to
put ourselves in the frame of the disk, and not in the frame of the observer. Thus, 
for an accurate CE, we must have the photon rates from lower energy range   
of $0.1-2$keV as well. That is why we choose to follow our procedure rather than using SIMPL 
model.

In several of the classes, we see that the photon numbers have two distinct states (see, \citet{skc02}
and references therein): one in which 
the count is lower (burst-off state designated as `h' in Tables 2 and 3) and the other in which 
the count is higher (burst-on state designated as `s' in Tables 2 and 3).

In Table 2, a comparison of the fitted parameters for a sample single 16s bin for all the classes 
are provided with an error at $90\%$ confidence level. 
Just to compare, we fit the spectrum of each class with the {\it diskbb} and {\it compST} model
to have the information of $\tau$ and $T_e$ of electron distribution inside the Compton cloud.  
The same sample 16s bin spectrum which were analyzed in Table 2 is taken for each class and fitted 
with the {\it diskbb} and {\it compST}  
model with $n_H=6.0 \times 10^{22}~ cm^{-2}$ hydrogen column for absorption and $0.5\%$ as the systematic error.
The best fitted parameters are given in Table 3 for comparison.
We find that in the cases where both models were satisfactory, the CE obtained were very close to each other. So our result appears to be
model independent.

\begin{table*}
\scriptsize{
\addtolength{\tabcolsep}{-2.5pt}
\begin{center}
\begin{tabular}{|c|c|c|c|c|c|c|c|c|c|}
\hline
\multicolumn{2}{|c|}{} & $T_{in}$ & $dbb_e$ & $\tilde{\chi}^2 $& Soft & Power law &  Hard & CE  & $\tilde{\chi}^2$ \\
\multicolumn{2}{|c|}{Class} &  &  & & Photon & index &  Photon &  &  \\
\multicolumn{2}{|c|}{} & (keV) & (keV) & (dofs) & (kphtns/s) &   & (kphtns/s) & (\%) & (dofs) \\
\hline
I($\phi$) &-&$1.67~^{+0.02}_{-0.02}$& 10.2 & 1.2(18) &$275.08~^{+8.77}_{-8.46}$ & $2.93~^{+0.27}_{-0.26}$ &$0.20~^{+0.04}_{-0.03}$ & $0.07~^{+0.02}_{-0.01}$ & 1.6(70)\\ 
\hline
II ($\delta$) &-&$1.65~^{+0.02}_{-0.01}$& 10.3 & 1.5(17) &$330.40~^{+9.48}_{-9.80}$ & $2.35~^{+0.24}_{-0.23}$ &$0.16~^{+0.03}_{-0.02}$ & $0.05~^{+0.01}_{-0.01}$ & 1.1(69)\\ 
\hline
III ($\gamma$) &-&$1.80~^{+0.01}_{-0.01}$& 10.4 & 1.2(17) & $511.13~^{+12.57}_{-12.22}$ & $2.92~^{+0.35}_{-0.31}$ & $0.31~^{+0.09}_{-0.07}$ & $0.06~^{+0.02}_{-0.01}$ & 1.5(58) \\ 
\hline
IV ($\omega$) &-&$1.94~^{+0.02}_{-0.02}$& 10.6 & 1.1(15) & $192.21~^{+7.80}_{-7.41}$ & $2.65~^{+0.31}_{-0.29}$ &$0.09~^{+0.02}_{-0.01}$ & $0.05~^{+0.01}_{-0.01}$ & 1.0(54)  \\ 
\hline
V ($\mu$) &-&$1.42~^{+0.02}_{-0.02}$& 7.0 & 1.5(10) & $468.53~^{+10.40}_{-10.02}$ & $2.33~^{+0.14}_{-0.14}$ &$0.63~^{+0.08}_{-0.07}$ & $0.13~^{+0.02}_{-0.02}$ & 1.0(66)\\ 
\hline
VI ($\nu$) &-&$1.75~^{+0.03}_{-0.03}$& 7.0 & 1.7(10) &$370.39~^{+14.15}_{-13.15}$ & $2.58~^{+0.12}_{-0.12}$ &$0.45~^{+0.04}_{-0.03}$ & $0.12~^{+0.02}_{-0.01}$ & 1.2(69)\\ 
\hline
VII ($\lambda$) &h& $0.93~^{+0.06}_{-0.05}$& 4.5 & 1.4(7) & $229.10~^{+28.94}_{-22.32}$ & $2.11~^{+0.12}_{-0.11}$ &$0.59~^{+0.07}_{-0.06}$ & $0.26~^{+0.08}_{-0.07}$ & 0.9(59) \\ 
\cline{2-10}
          &s& $1.99~^{+0.02}_{-0.02}$& 9.0 & 1.1(14) & $437.22~^{+11.11}_{-10.62}$ & $2.68~^{+0.22}_{-0.21}$ &$0.19~^{+0.3}_{-0.2}$ & $0.04~^{+0.01}_{-0.01}$ & 1.2(69) \\ 
\hline
VIII ($\kappa$) &h& $1.15~^{+0.06}_{-0.05}$& 6.5 & 1.4(10) & $156.03~^{+17.10}_{-13.90}$ & $2.06~^{+0.15}_{-0.15}$ &$0.39~^{+0.06}_{-0.05}$ & $0.25~^{+0.08}_{-0.07}$ & 1.2(60)  \\ 
\cline{2-10}
          &s& $1.90~^{+0.03}_{-0.03}$& 8.0 & 0.9(11) & $432.72~^{+13.95}_{-13.15}$ & $2.98~^{+0.24}_{-0.23}$ &$0.21~^{+0.03}_{-0.03}$ & $0.05~^{+0.01}_{-0.01}$ & 1.5(51)  \\
\hline
IX ($\rho$) &h&$1.38~^{+0.03}_{-0.03}$& 5.7 & 1.6(8) & $382.33~^{+18.42}_{-16.73}$ & $2.50~^{+0.10}_{-0.10}$ &$0.95~^{+0.09}_{-0.08}$ & $0.25~^{+0.04}_{-0.03}$ & 1.5(52) \\ 
\cline{2-10}
       &s&$1.84~^{+0.02}_{-0.02}$& 8.0 & 1.7(11) & $419.92~^{+11.55}_{-10.97}$ & $2.33~^{+0.16}_{-0.15}$ &$0.34~^{+0.04}_{-0.03}$ & $0.08~^{+0.01}_{-0.01}$ & 1.0(55)  \\ 
\hline
X ($\xi$) &h& $1.14~^{+0.05}_{-0.04}$& 5.5 & 0.8(11) & $162.63~^{+21.83}_{-18.49}$ & $2.82~^{+0.19}_{-0.18}$ &$0.33~^{+0.11}_{-0.09}$ & $0.20~^{+0.08}_{-0.10}$ & 1.1(52)   \\  
\cline{2-10}
        &s& $1.39~^{+0.02}_{-0.02}$& 9.5 & 1.3(13) & $106.38~^{+5.83}_{-5.50}$ & $2.41~^{+0.48}_{-0.44}$ &$0.07~^{+0.09}_{-0.04}$ & $0.06~^{+0.02}_{-0.01}$ & 1.1(55)\\ 
\hline
XI ($\beta$) &h& $1.04~^{+0.03}_{-0.03}$& 5.2 & 1.0(10) & $320.15~^{+20.78}_{-18.15}$ & $2.14~^{+0.06}_{-0.06}$ &$0.87~^{+0.05}_{-0.04}$ & $0.27~^{+0.04}_{-0.04}$ & 1.1(69)   \\ 
\cline{2-10}
        &s& $1.45~^{+0.02}_{-0.02}$& 7.0 & 1.5(12) & $354.87~^{+12.38}_{-11.58}$ & $2.69~^{+0.18}_{-0.17}$ &$0.32~^{+0.05}_{-0.04}$ & $0.14~^{+0.03}_{-0.02}$ & 1.6(60)\\ 
\hline
XII ($\alpha$) &-&$1.21~^{+0.08}_{-0.07}$& 4.5 & 1.1(8) &$261.36~^{+38.46}_{-28.74}$ & $2.07~^{+0.06}_{-0.06}$ &$0.81~^{+0.04}_{-0.04}$ & $0.31~^{+0.08}_{-0.07}$ & 1.0(65)  \\ 
\hline
XIII ($\theta$) &h&$1.39~^{+0.04}_{-0.04}$& 6.0 & 1.3(10) &$416.84~^{+25.73}_{-22.85}$ & $2.50~^{+0.09}_{-0.09}$ &$1.15~^{+0.09}_{-0.08}$ & $0.28~^{+0.05}_{-0.04}$ & 1.6(58)  \\ 
\cline{2-10}
         &s&$1.32~^{+0.03}_{-0.03}$& 5.5 & 1.6(10) & $388.95~^{+21.46}_{-19.19}$ & $3.25~^{+0.16}_{-0.15}$ &$0.65~^{+0.08}_{-0.07}$ & $0.17~^{+0.03}_{-0.03}$ & 1.5(63)  \\ 
\hline
XIV ($\chi_2$) &-&$1.20~^{+0.12}_{-0.09}$& 4.5 & 1.1(6) &$126.80~^{+27.74}_{-18.46}$ & $1.97~^{+0.08}_{-0.07}$ &$0.57~^{+0.03}_{-0.03}$ & $0.45~^{+0.17}_{-0.12}$ & 1.6(60)  \\ 
\hline
XV ($\chi_4$) &-&$1.18~^{+0.11}_{-0.09}$& 4.5 & 1.2(6) &$121.04~^{+25.87}_{-17.37}$ & $1.95~^{+0.08}_{-0.08}$ &$0.52~^{+0.03}_{-0.02}$ & $0.43~^{+0.16}_{-0.12}$ & 1.4(60)  \\ 
\hline
XVI ($\chi_1$) &-&$1.16~^{+0.05}_{-0.05}$& 4.5 & 1.4(6) & $599.39~^{+58.18}_{-47.51}$ & $2.74~^{+0.06}_{-0.06}$ &$4.62~^{+0.12}_{-0.11}$ & $0.77~^{+0.08}_{-0.07}$ & 1.7(57)  \\ 
\hline
XVII ($\chi_3$) &-&$1.19~^{+0.03}_{-0.03}$& 5.0 & 1.3(6) & $598.83~^{+31.99}_{-28.57}$ & $2.70~^{+0.06}_{-0.06}$ &$5.30~^{+0.09}_{-0.09}$ & $0.88~^{+0.10}_{-0.09}$ & 1.3(61)\\ 
\hline
\end{tabular}
\end{center}
\caption{Parameters for the spectral fits of sample single 16s bin with {\it diskbb} 
plus {\it powerlaw} models for all the variability classes. $T_{in}$ is the black body temperature 
obtained from fitting. $dbb_e$ is the upper limit of the disk blackbody spectrum. The column 
containing the soft photon rate contains blackbody photons in the $0.1-dbb_e$ keV. 
The column `power-law' contains the power-law index $\alpha$ obtained from fitting.
The column `hard photons' contains the rate at which Comptonized photons are emitted in the
range $3 \times T_{in} - 40$ keV. CE is the Comptonized efficiency.}}
\end{table*}

\begin{table*}
\scriptsize{
\addtolength{\tabcolsep}{-4.5pt}
\begin{center}
\begin{tabular}{|c|c|c|c|c|c|c|c|c|c|c|}
\hline
\multicolumn{2}{|c|}{}& $T_{in}$ & $dbb_e$ &$\tilde{\chi}^2$ & Soft & $T_e$ & $\tau$ &  Hard & CE  & $\tilde{\chi}^2$ \\
\multicolumn{2}{|c|}{Class} &  &  & &  Photon & & & Photon &  & \\
\multicolumn{2}{|c|}{}& (keV) & (keV) & (dofs) & (kphotons/sec) & keV &  & (kphotons/s) & (\%) & (dofs) \\
\hline
I &-&$1.67~^{+0.02}_{-0.02}$& 10.2 & 1.2(18)& $275.08~^{+8.77}_{-8.46}$ &-&-&-&-&- \\ 
\hline
II &-&$1.65~^{+0.02}_{-0.01}$& 10.3 & 1.5(17) & $330.04~^{+9.48}_{-9.80}$ &-&-&-&-&- \\ 
\hline
III &-&$1.80~^{+0.01}_{-0.01}$& 10.4 & 1.2(17) & $511.13~^{+12.57}_{-12.22}$ &-&-&-&-&- \\ 
\hline
IV &-&$1.94~^{+0.02}_{-0.02}$& 10.6 & 1.1(15) & $192.21~^{+7.80}_{-7.41}$ &-&-&-&-&-   \\ 
\hline
V &-&$1.42~^{+0.02}_{-0.02}$& 7.0& 1.5(10) & $468.53~^{+10.40}_{-10.02}$ & $4.72~^{+0.98}_{-0.66}$ & $14.40~^{+4.62}_{-4.46}$ & $0.61~^{+0.08}_{-0.07}$ & $0.13~^{+0.02}_{-0.02}$ & 1.0(61)  \\ 
\hline
VI &-&$1.75~^{+0.03}_{-0.03}$& 7.0 & 1.7(10) & $370.39~^{+14.15}_{-13.15}$ &-&-&-&-&- \\ 
\hline
VII &h& $0.93~^{+0.06}_{-0.05}$& 4.5 & 1.4(7) & $229.10~^{+28.94}_{-22.32}$ & $4.93~^{+1.08}_{-0.68}$ & $12.43~^{+2.75}_{-2.25}$ & $0.52^{+0.07}_{-0.06}$ & $0.22~^{+0.08}_{-0.07}$ & 1.5(60) \\ 
\cline{2-11}
          &s & $1.99~^{+0.02}_{-0.02}$& 9.0 & 1.1(14) & $437.22~^{+11.11}_{-10.62}$ &-&-&-&-&-  \\ 
\hline
VIII &h& $1.15~^{+0.06}_{-0.05}$& 6.5 & 1.4(10) & $156.03~^{+17.10}_{-13.90}$ & $6.41~^{+2.56}_{-1.25}$ & $9.41~^{+2.88}_{-2.39}$ & $0.41~^{+0.06}_{-0.05}$ & $0.26~^{+0.08}_{-0.07}$ & 0.9(58)  \\ 
\cline{2-11}
          &s& $1.90~^{+0.03}_{-0.03}$& 8.0 & 0.9(11) & $437.72~^{+13.95}_{-13.15}$ &-&-&-&-&-   \\
\hline
IX &h&$1.38~^{+0.03}_{-0.03}$& 5.7 & 1.6(8) & $382.33~^{+18.42}_{-16.73}$ & $3.89~^{+0.46}_{-0.29}$ & $30.93~^{+4.36}_{-4.36}$ & $0.75~^{+0.09}_{-0.08}$ & $0.20~^{+0.04}_{-0.03}$ & 1.4(61) \\ 
\cline{2-11}
       &s&$1.84~^{+0.02}_{-0.02}$& 8.0 & 1.7(11) &$419.92~^{+11.55}_{-10.97}$ &-&-&-&-&-  \\ 
\hline
X &h& $1.14~^{+0.05}_{-0.04}$& 5.5 & 0.8(11) & $162.63~^{+21.83}_{-18.49}$ & $3.19~^{+1.15}_{-0.59}$ & $14.36~^{+27.70}_{-5.72}$ & $0.29~^{+0.09}_{-0.08}$ & $0.17~^{+0.05}_{-0.06}$ &  0.8(58) \\ 
\cline{2-11}
         &s& $1.39~^{+0.02}_{-0.02}$& 9.5 & 1.3(13) & $106.38~^{+5.83}_{-5.50}$ &-&-&-&-&- \\     
\hline
XI  &h& $1.04~^{+0.03}_{-0.03}$& 5.2 & 1.0(10) &$320.15~^{+20.78}_{-18.15}$ & $4.61~^{+0.82}_{-0.56}$ & $11.07~^{+2.88}_{-2.10}$ & $0.57~^{+0.05}_{-0.04}$ & $0.18~^{+0.04}_{-0.04}$ & 0.9(56)   \\ 
\cline{2-11}
         &s& $1.45~^{+0.02}_{-0.02}$& 7.0& 1.5(12) & $354.87~^{+12.38}_{-11.58}$ &-&-&-&-&- \\ 
\hline
XII &-&$1.21~^{+0.08}_{-0.07}$& 4.5 & 1.1(8) &$261.36~^{+38.46}_{-28.74}$ & $5.04~^{+0.51}_{-0.41}$ & $11.91~^{+1.54}_{-1.32}$ & $0.71~^{+0.04}_{-0.04}$ & $0.27~^{+0.08}_{-0.07}$ & 1.4(58) \\ 
\hline
XIII &h&$1.39~^{+0.04}_{-0.04}$& 6.0 & 1.3(10) & $416.84~^{+25.73}_{-22.85}$ & $4.13~^{+0.67}_{-0.47}$ & $14.41~^{+1.49}_{-4.10}$ &$0.96~^{+0.09}_{-0.08}$ & $0.43~^{+0.07}_{-0.07}$ & 1.3(64)  \\ 
\cline{2-11}
         &s& $1.32~^{+0.03}_{-0.03}$ & 5.5 & 1.6(10) & $388.95~^{+21.46}_{-19.19}$ &-&-&-&-&-  \\ 
\hline
XIV &-&$1.20~^{+0.12}_{-0.09}$& 4.5 & 1.1(6) & $126.80~^{+27.74}_{-18.46}$ & $8.92~^{+1.95}_{-2.30}$ & $6.80~^{+1.93}_{-2.64}$ & $0.52^{+0.03}_{-0.03}$ & $0.41~^{+0.17}_{-0.12}$ & 1.5(61)  \\ 
\hline
XV &-&$1.18~^{+0.11}_{-0.09}$& 4.5 & 1.2(6)& $121.04~^{+25.87}_{-17.37}$ & $6.62~^{+2.74}_{-1.23}$ & $8.79~^{+2.22}_{-2.15}$ &$0.51~^{+0.03}_{-0.02}$ & $0.42~^{+0.16}_{-0.12}$ & 0.90(60)  \\ 
\hline
XVI &-&$1.16~^{+0.05}_{-0.05}$& 4.5 & 1.4(6) & $599.39~^{+58.18}_{-47.51}$ & $4.64~^{+0.82}_{-0.55}$ & $8.50~^{+1.58}_{-1.40}$ & $4.10~^{+0.12}_{-0.11}$ & $0.68~^{+0.08}_{-0.07}$ & 1.3(62)  \\ 
\hline
XVII &-&$1.19~^{+0.03}_{-0.03}$& 5.0 & 1.3(6) & $598.83~^{+31.99}_{-28.57}$ & $6.62~^{+2.74}_{-1.23}$ & $8.79~^{+2.22}_{-2.15}$ & $5.20~^{+0.09}_{-0.09}$ & $0.87~^{+0.10}_{-0.09}$ & 1.2(60)  \\ 
\hline
\end{tabular}
\end{center}
\caption{Same data as in Table 2, analyzed with the {\it diskbb} and {\it comptST} model.
Most of the columns have the same meaning as in Table 2. However, the columns containing 
the electron temperature $T_e$ and the optical depth $\tau$ have been added.
}}
\end{table*}

\section{Results}

We discuss below the main results for each class. The Observational IDs are given in Table 1.
So far, we showed that the value of CE, averaged over a class, varies from one variability class to another.
We now present the results of the dynamical analysis of the light curves of all the variability classes.  
While choosing the data duration for a given class, the following considerations have been made:
If the count rate has no obvious high and low count features, we use the data of $500$s. If the count rates
have some `repetitive' behavior, we use the data of $500$, $1000$, $1500$s or even $2000$s, which ever is 
bigger so that at least one full cycle in included in the chunk of the data. 
However, in the latter cases, while computing the average value of CE, we use the data from a full cycle only. 

In each of Figs. 2(a-q) we show four panels. The top panel is the variation of the photons
with time. The second panel is the variation of 
log(power) of the dynamical Power Density Spectra (PDS) which may show presence or absence of quasi-periodic
variations or QPOs. The third panel is the dynamic energy spectrum which shows whether the variability class
is dominated by power-law photons or black-body photons. Finally, the bottom panel shows the 
variation of Comptonizing Efficiency CE calculated using $16$ seconds of binned data. The error bar is provided
at the $90\%$ confidence level.

   \begin{figure*}
   \centering
{
   \includegraphics[angle=270,width=9.50cm]{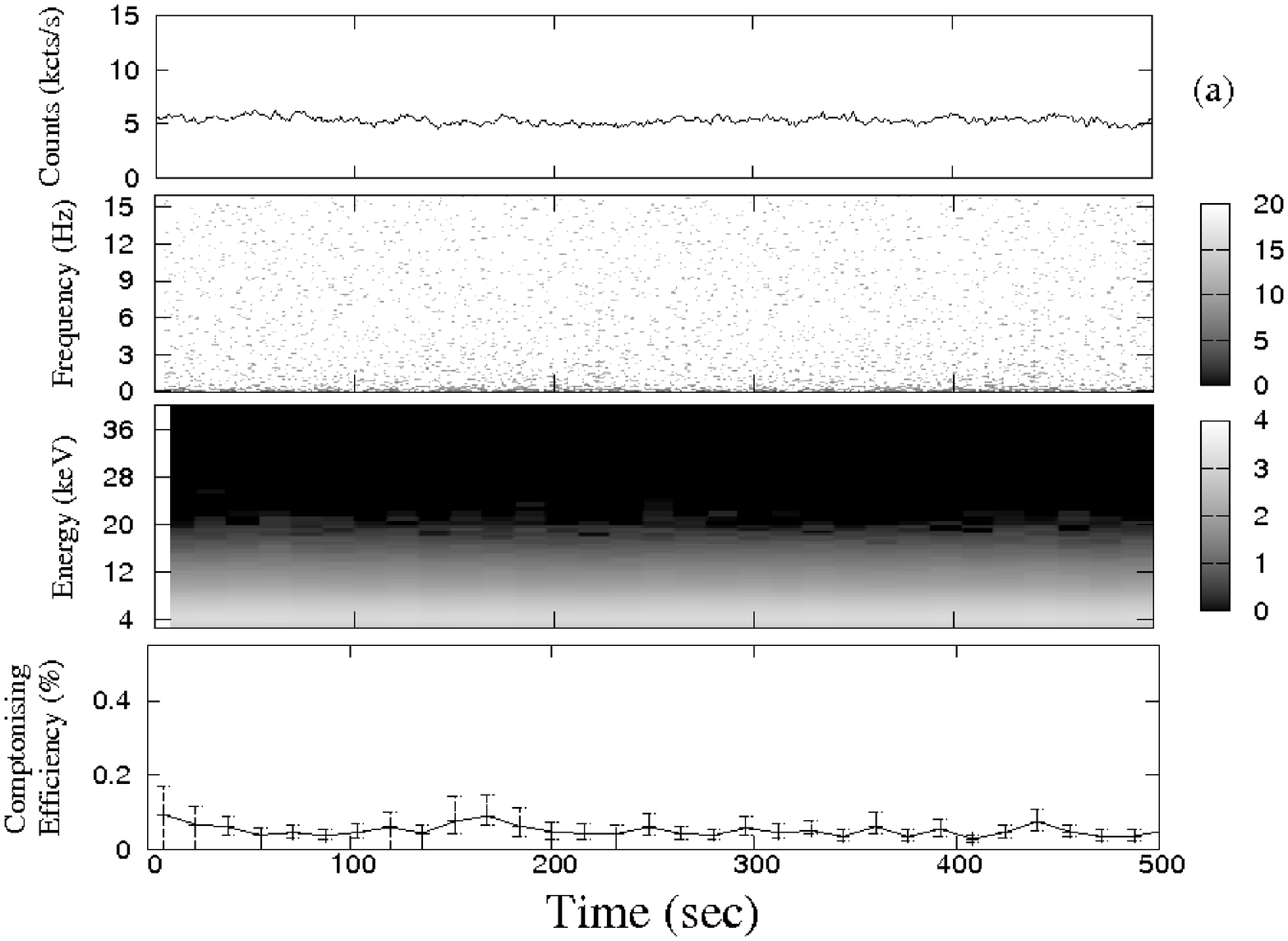}
   \includegraphics[angle=270,width=9.50cm]{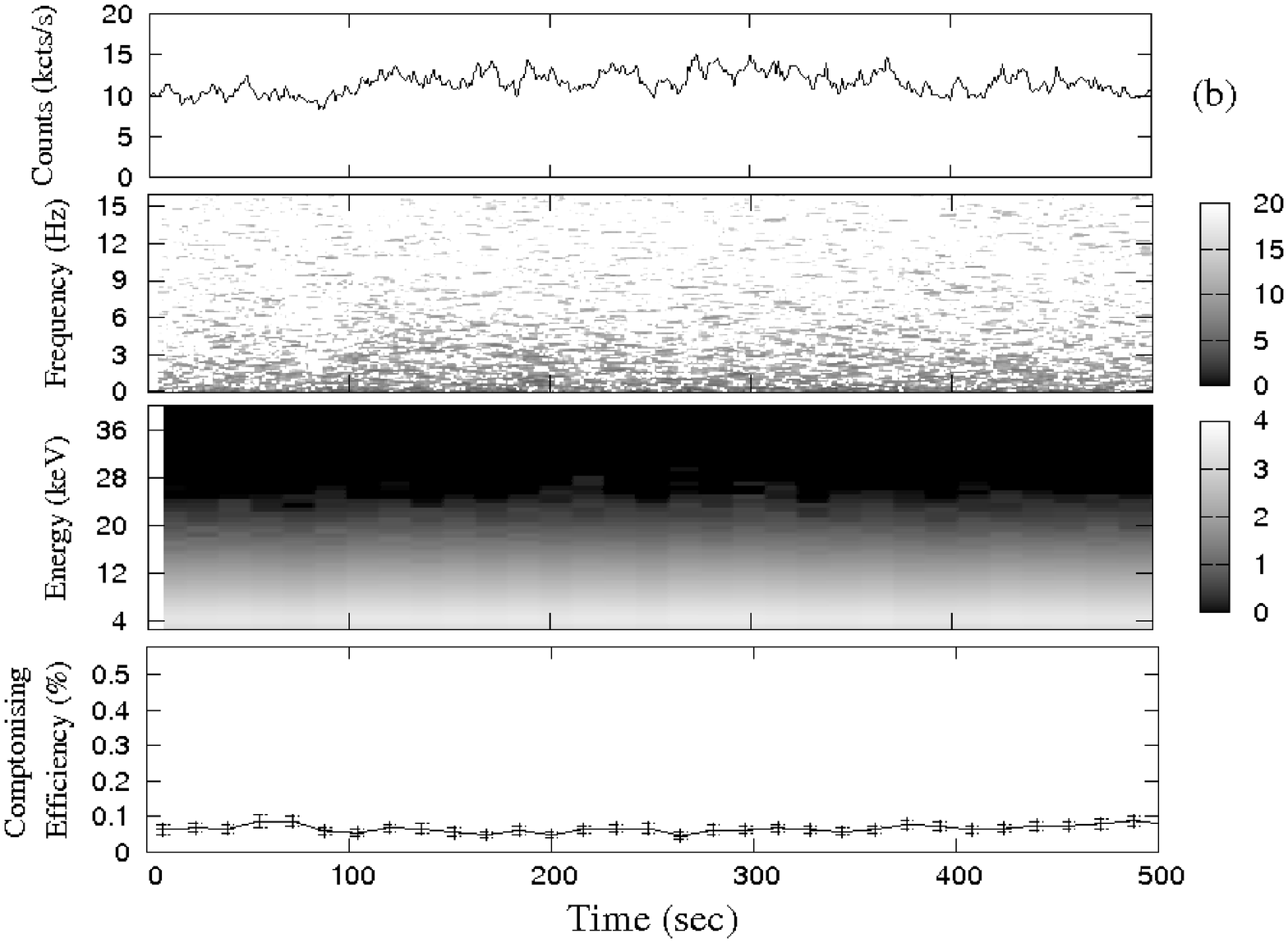} 
   \includegraphics[angle=270,width=9.50cm]{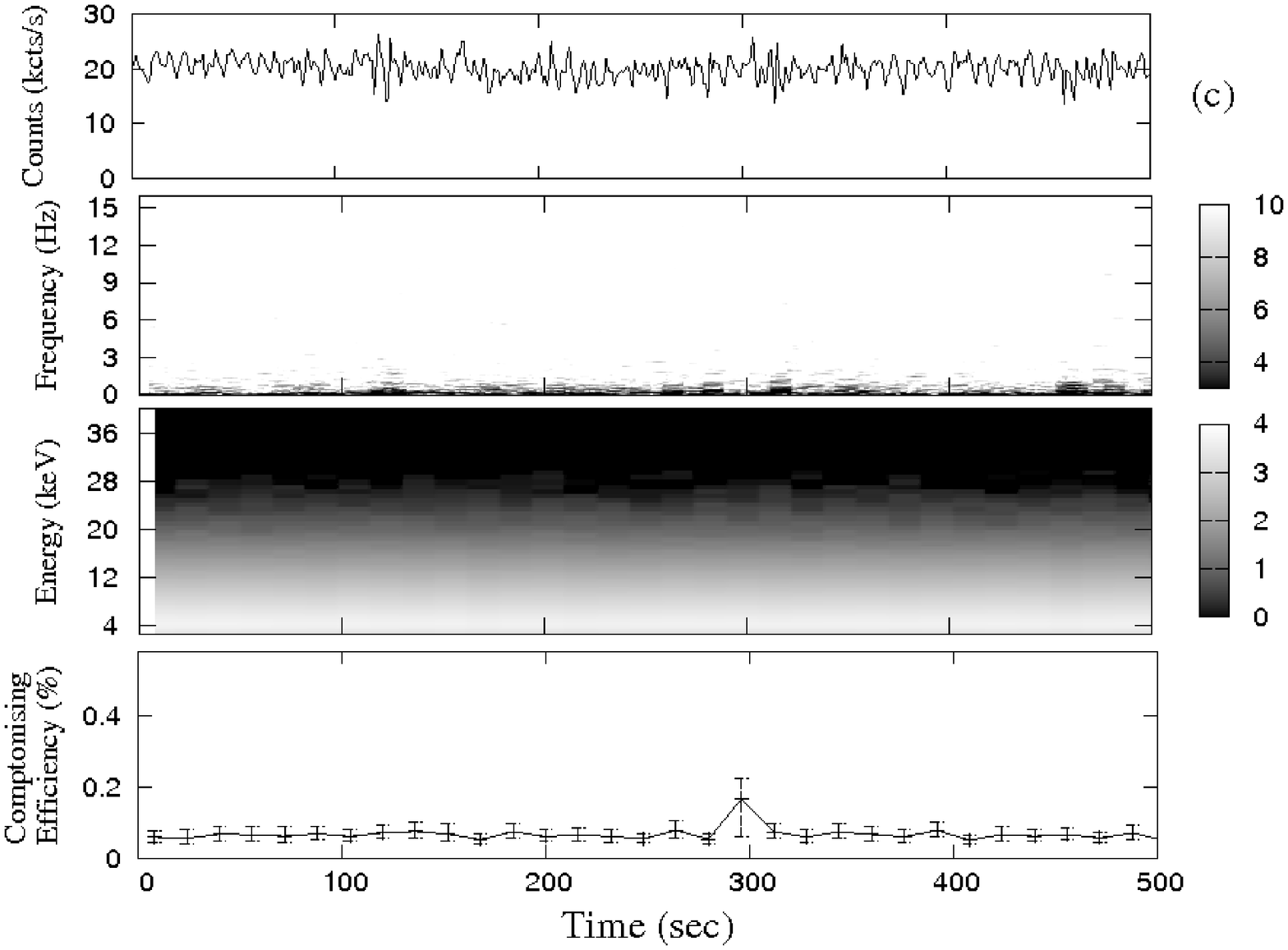}
}
   \label{fig3}
   \end{figure*}

   \begin{figure*}
   \centering
{
   \includegraphics[angle=270,width=9.50cm]{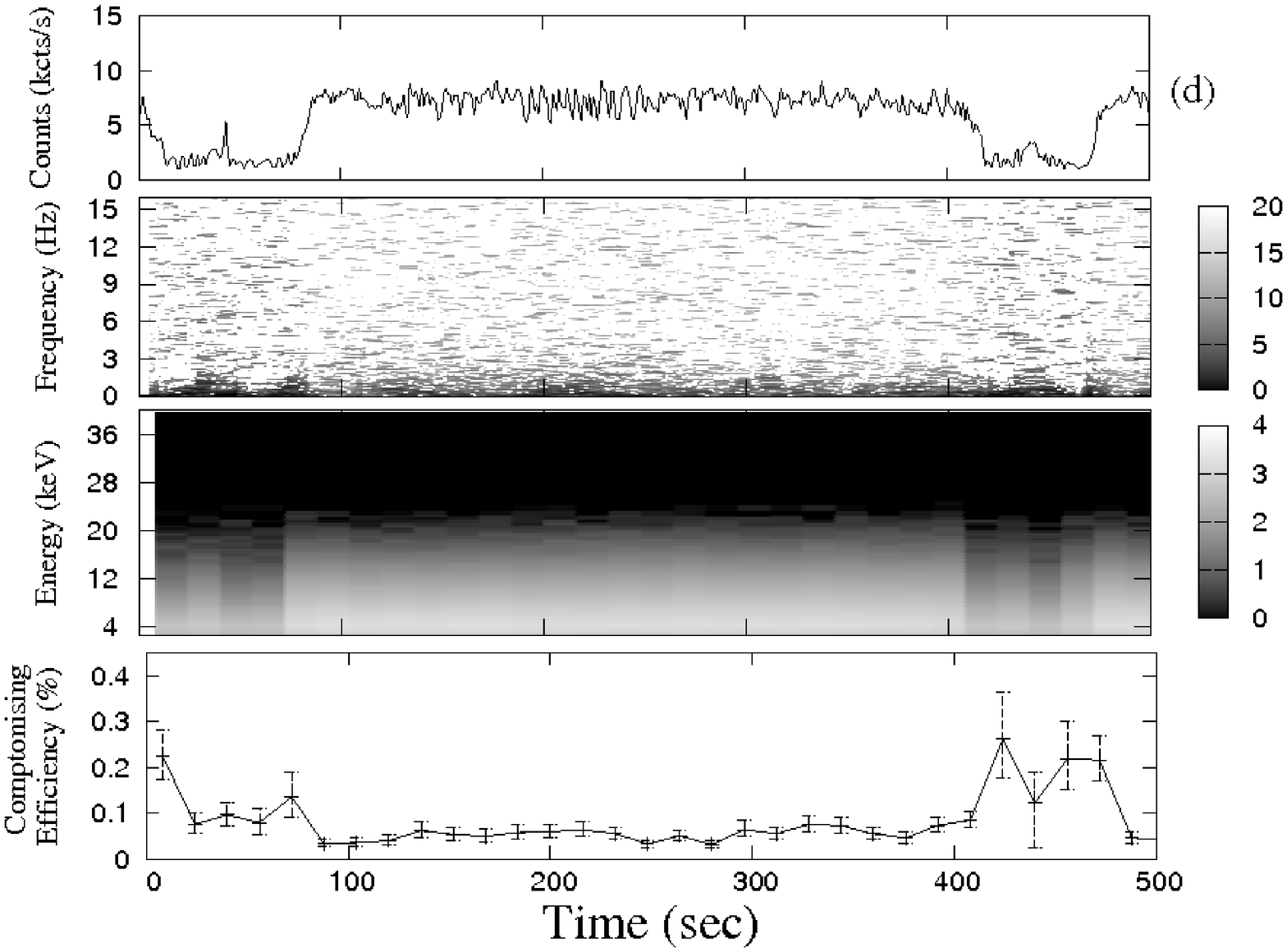}
   \includegraphics[angle=270,width=9.50cm]{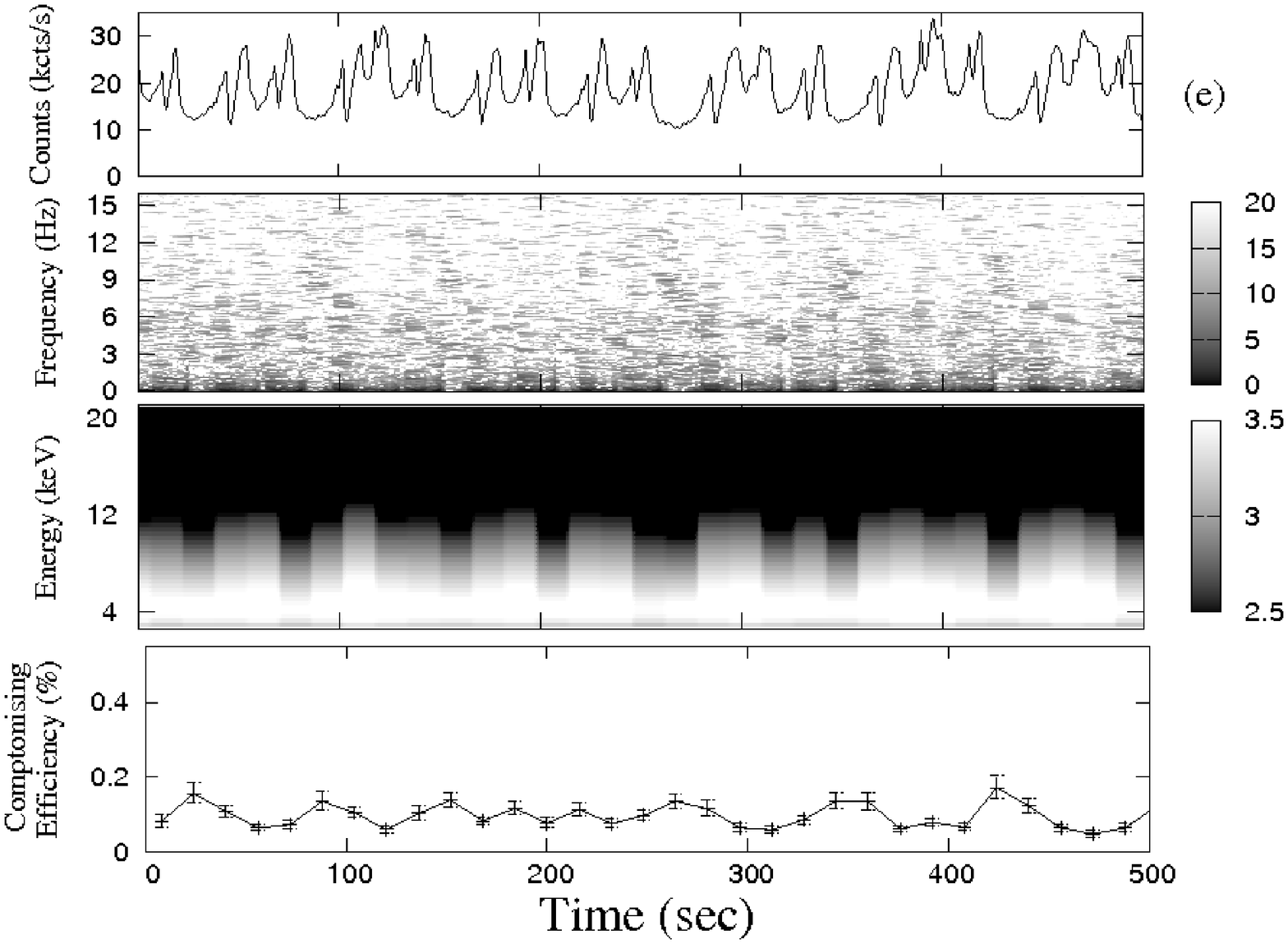}
   \includegraphics[angle=270,width=9.50cm]{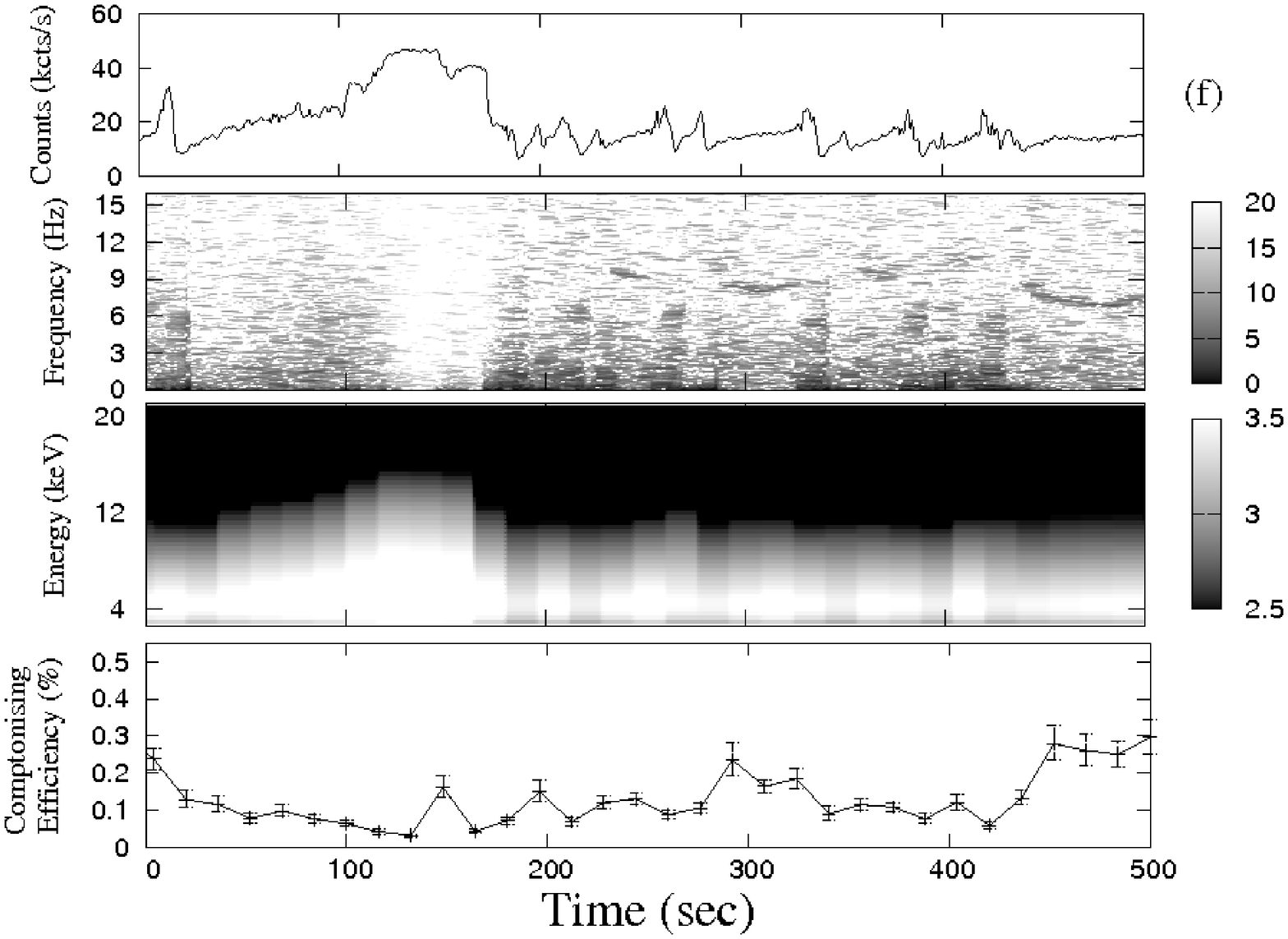}
}
   \label{fig3}
   \end{figure*}

   \begin{figure*}
   \centering
{
   \includegraphics[angle=270,width=9.50cm]{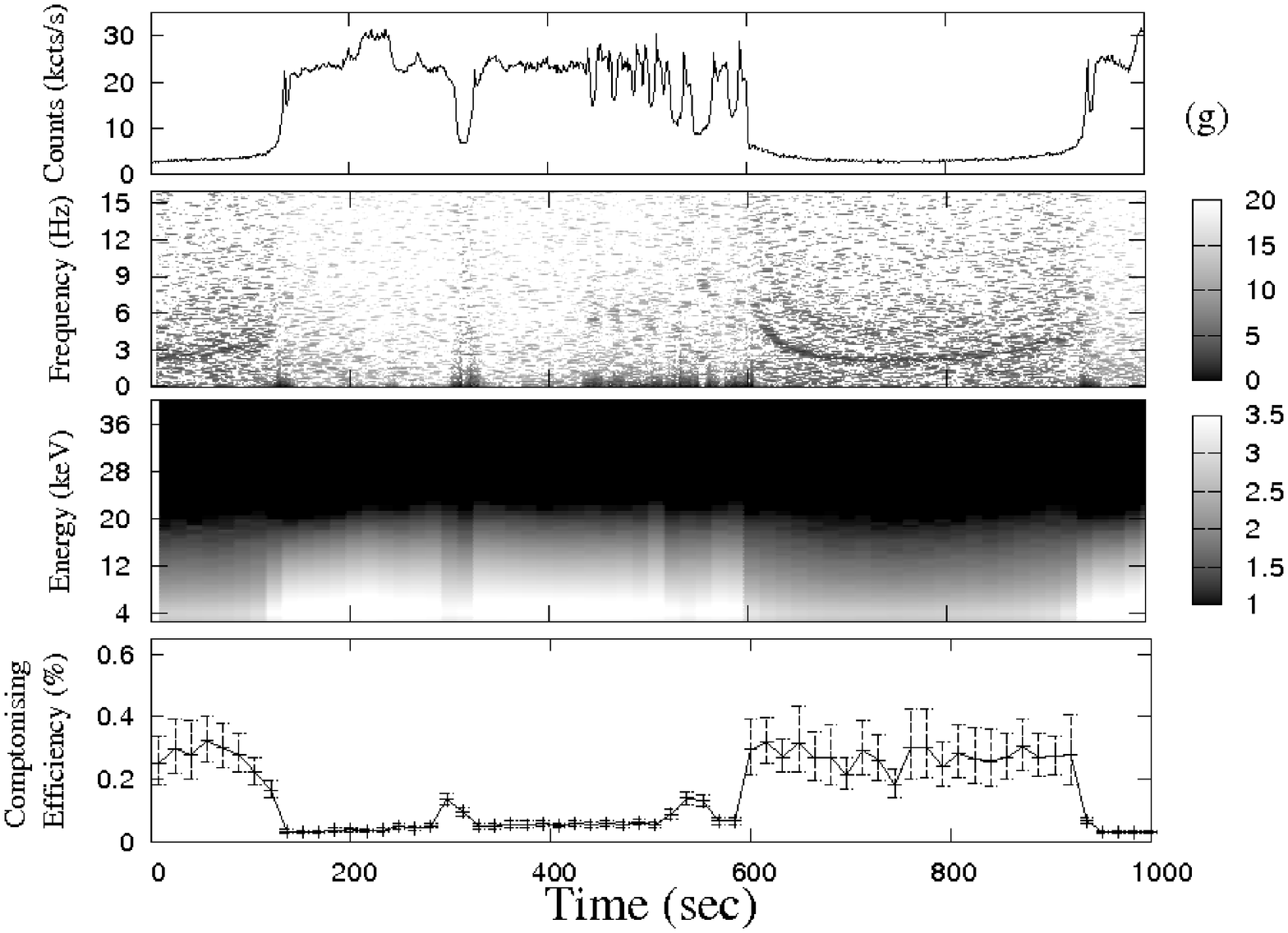}
   \includegraphics[angle=270,width=9.50cm]{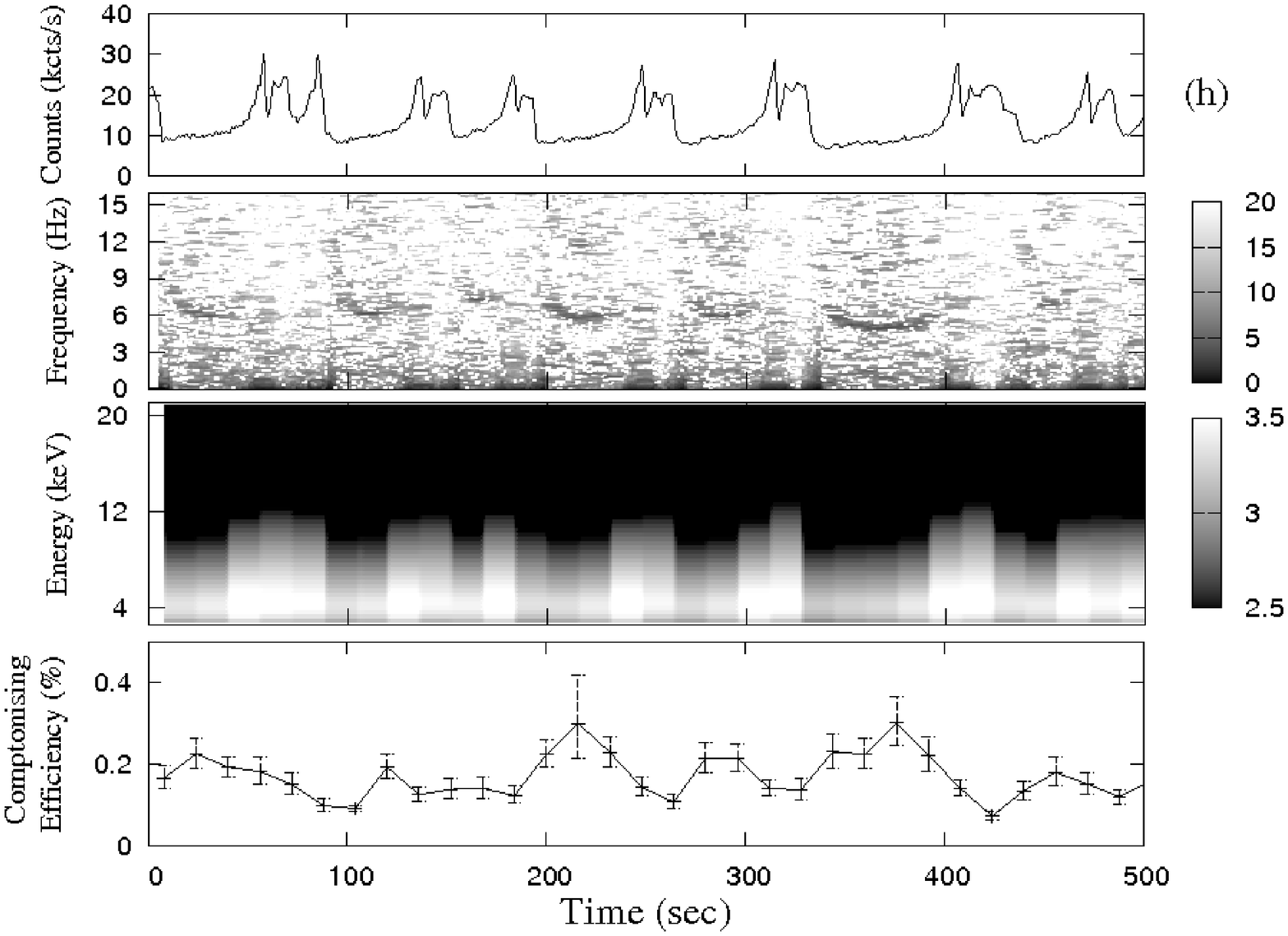}
  \includegraphics[angle=270,width=9.50cm]{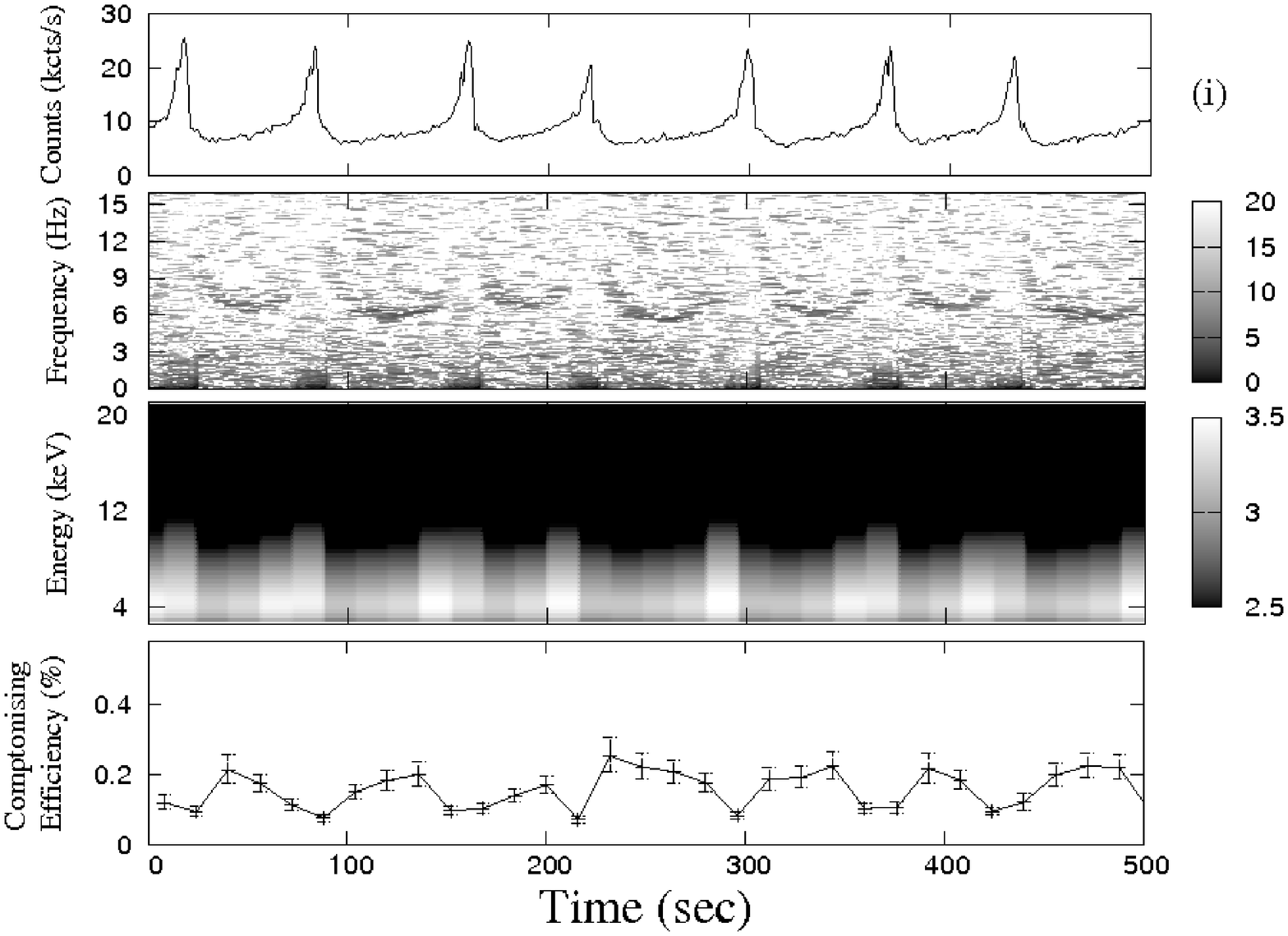}
}
   \label{fig3}
   \end{figure*}

   \begin{figure*}
   \centering
{
  \includegraphics[angle=270,width=9.50cm]{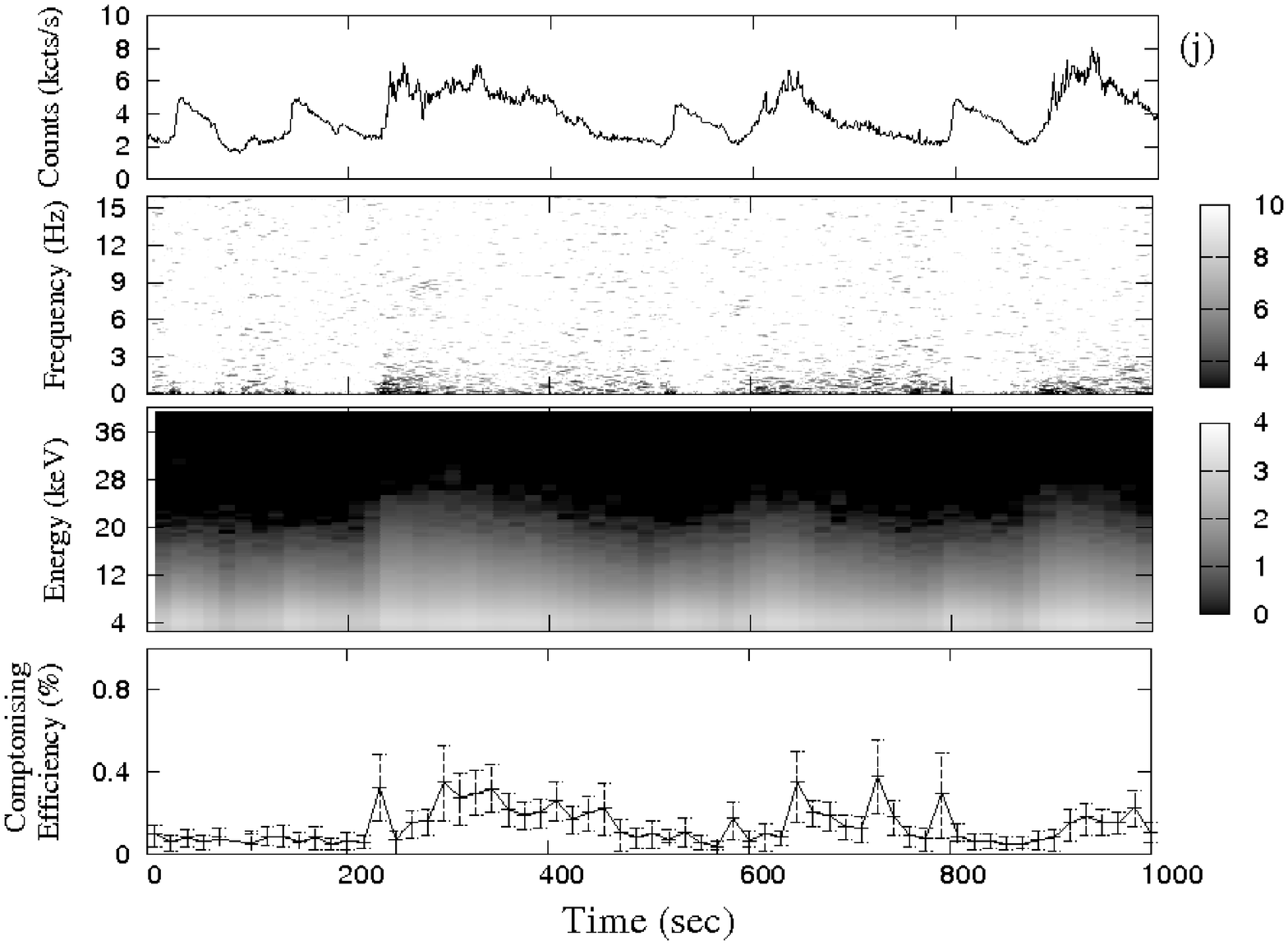}
   \includegraphics[angle=270,width=9.50cm]{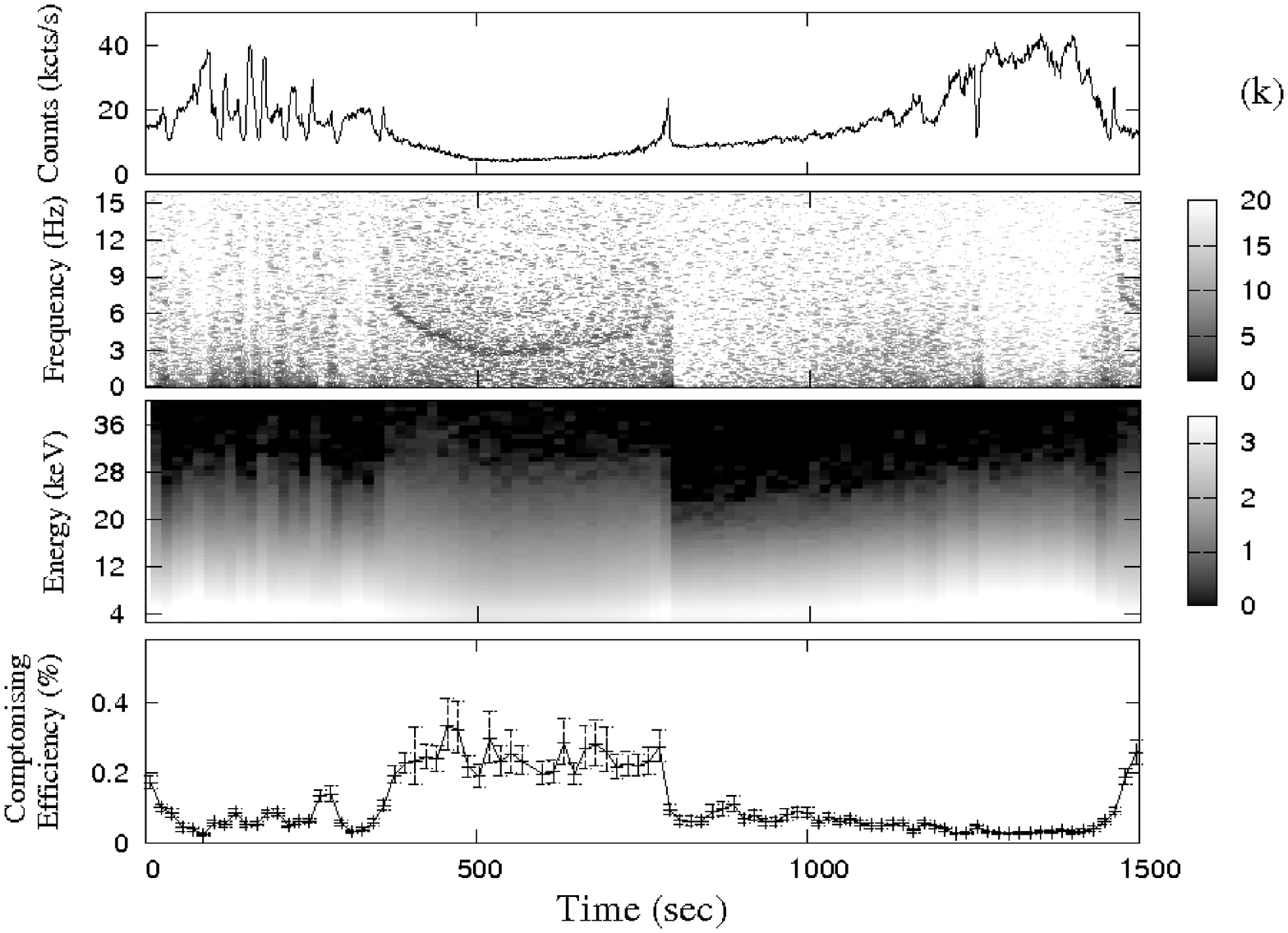}
   \includegraphics[angle=270,width=9.50cm]{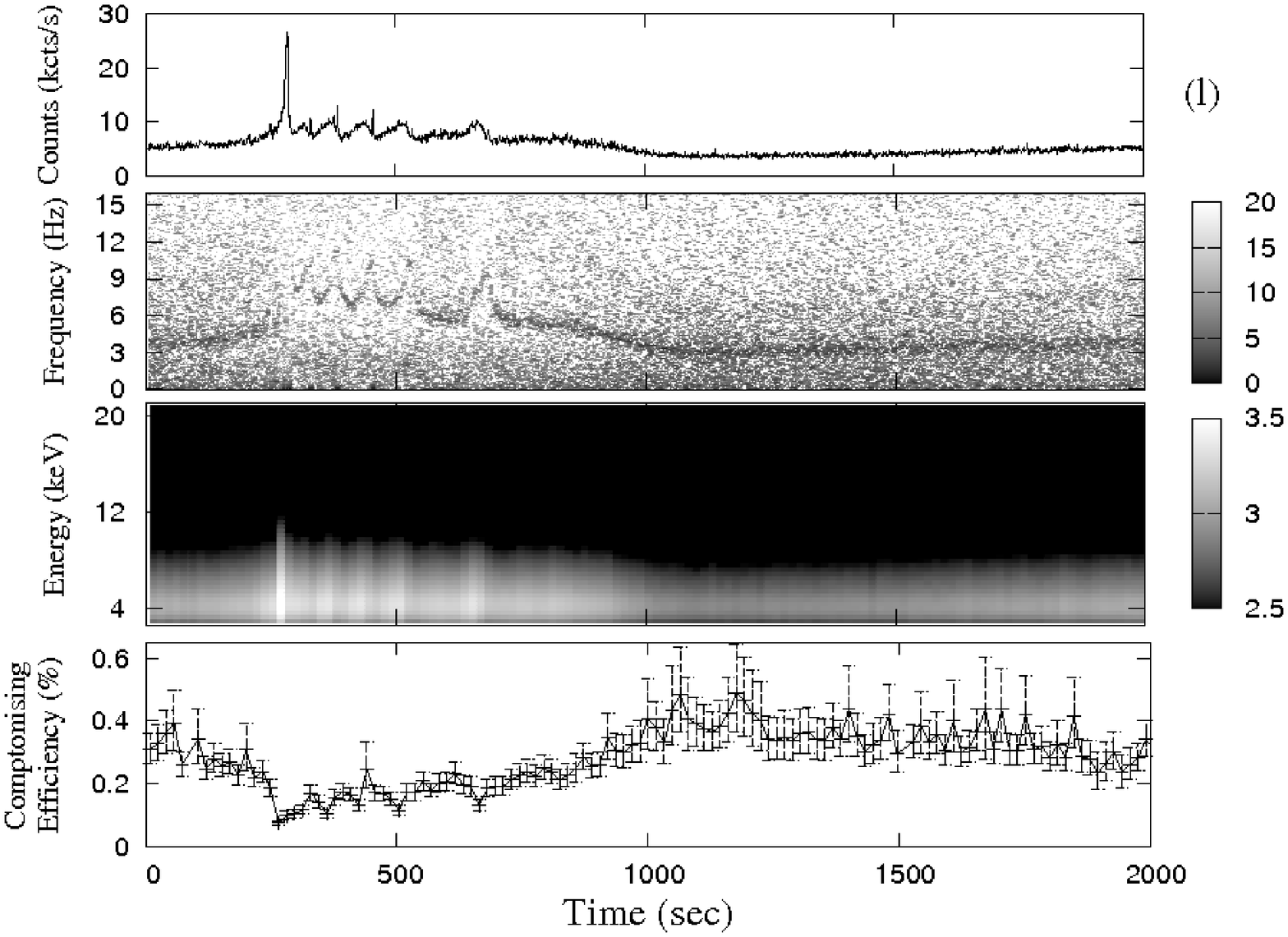}
}
   \label{fig3}
   \end{figure*}


   \begin{figure*}
   \centering
{
   \includegraphics[angle=270,width=9.50cm]{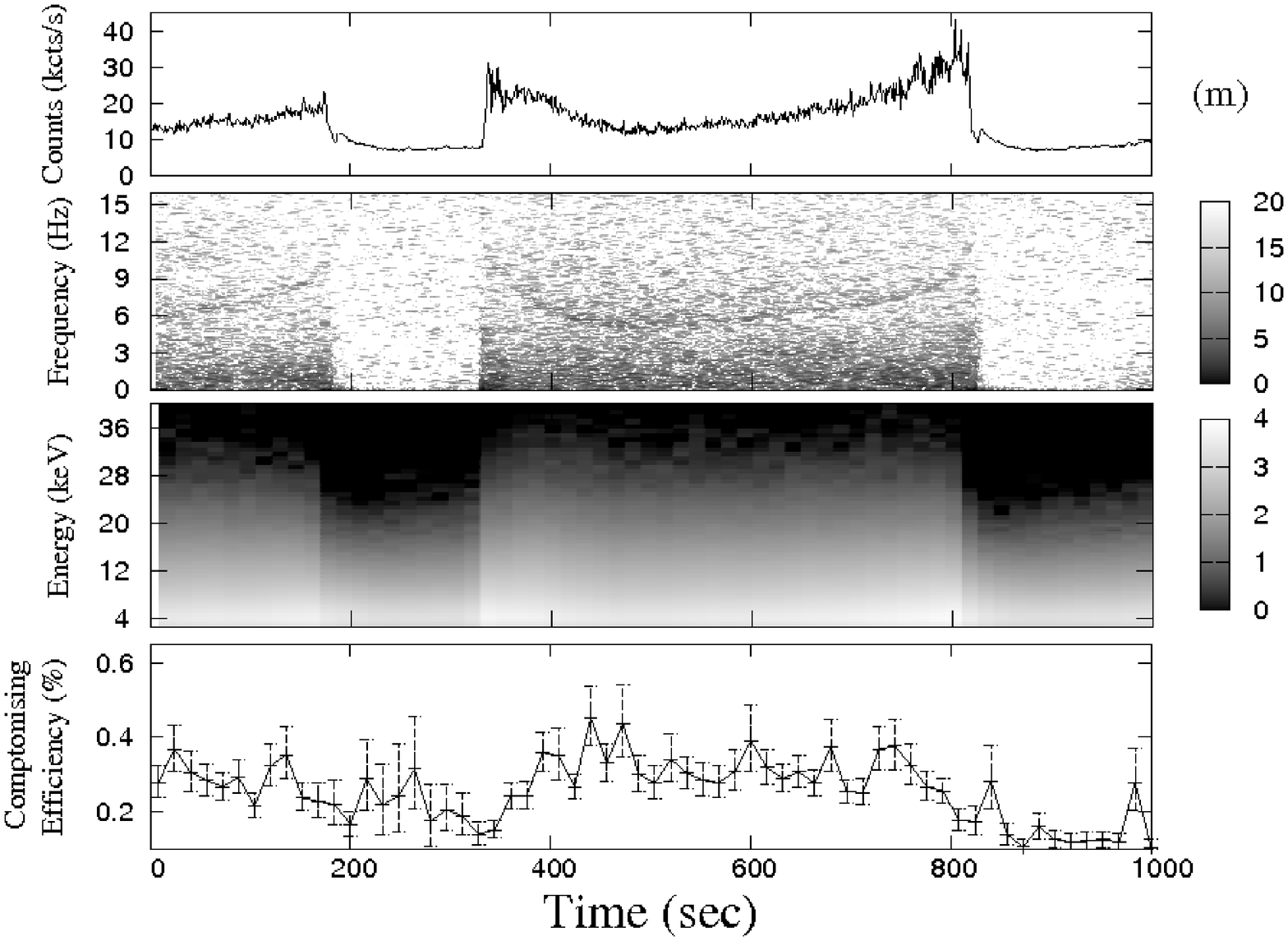}
   \includegraphics[angle=270,width=9.50cm]{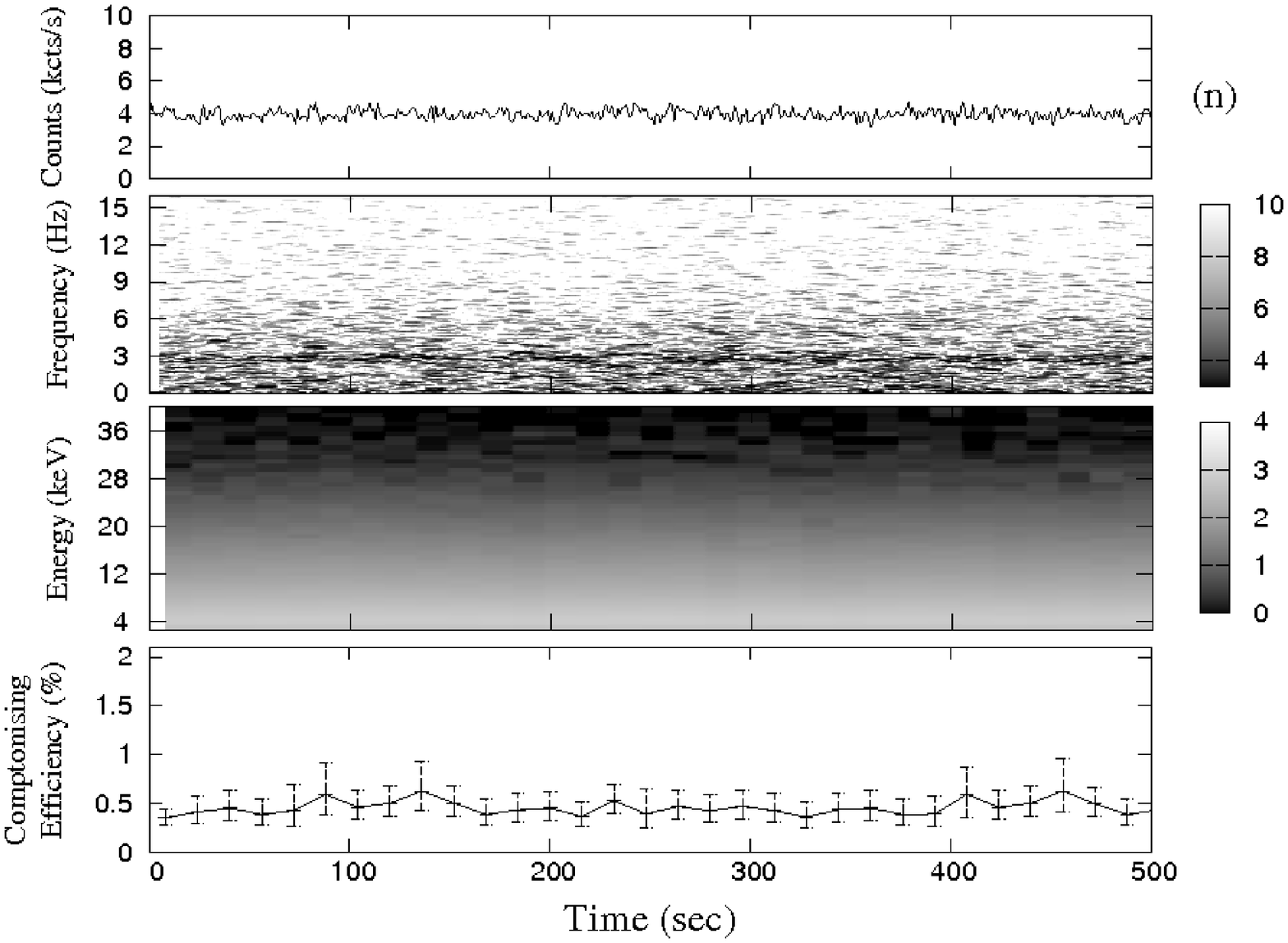}
   \includegraphics[angle=270,width=9.50cm]{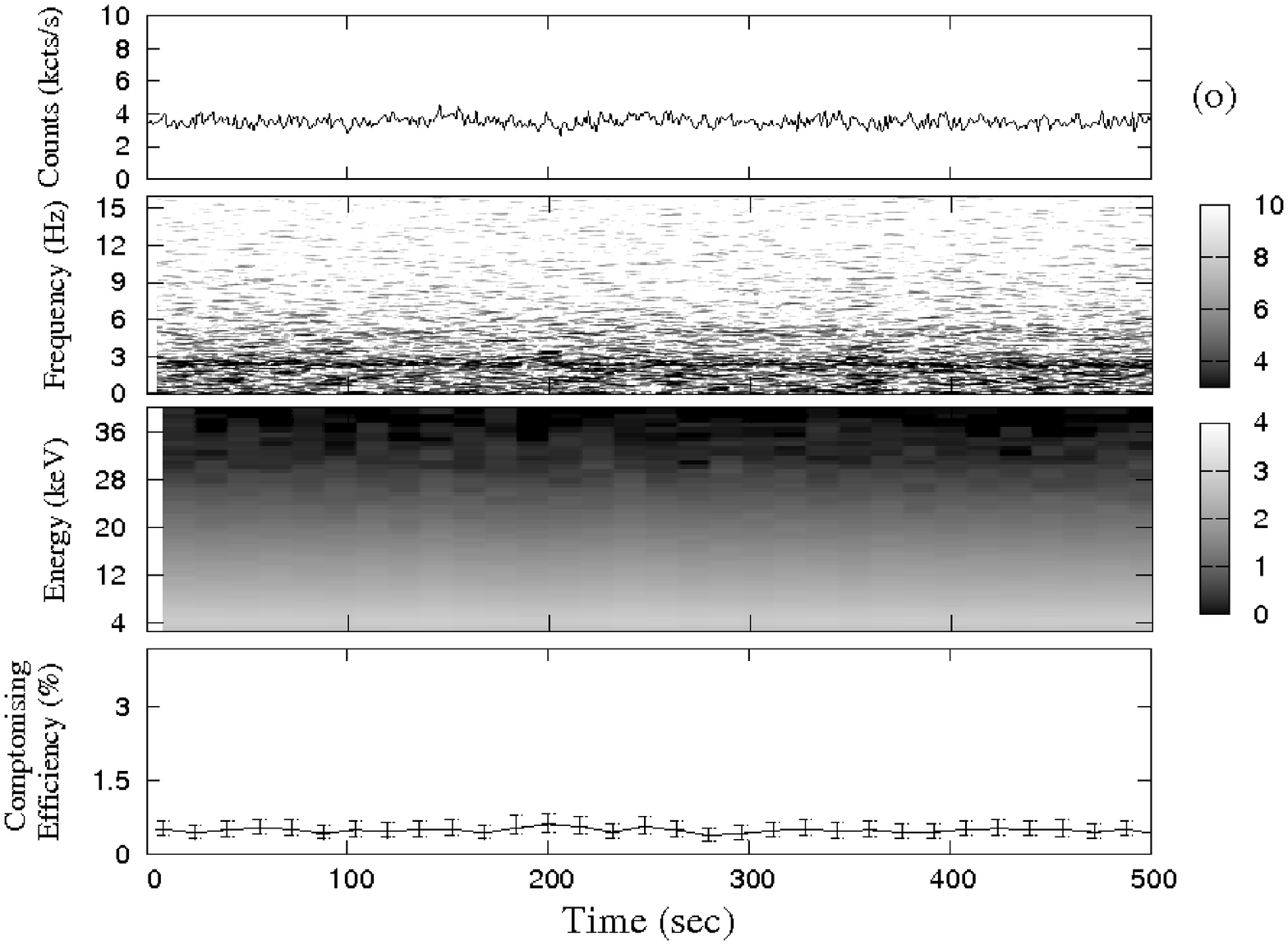}
}
   \label{fig3}
   \end{figure*}

   \begin{figure*}
   \centering
{
   \includegraphics[angle=270,width=9.50cm]{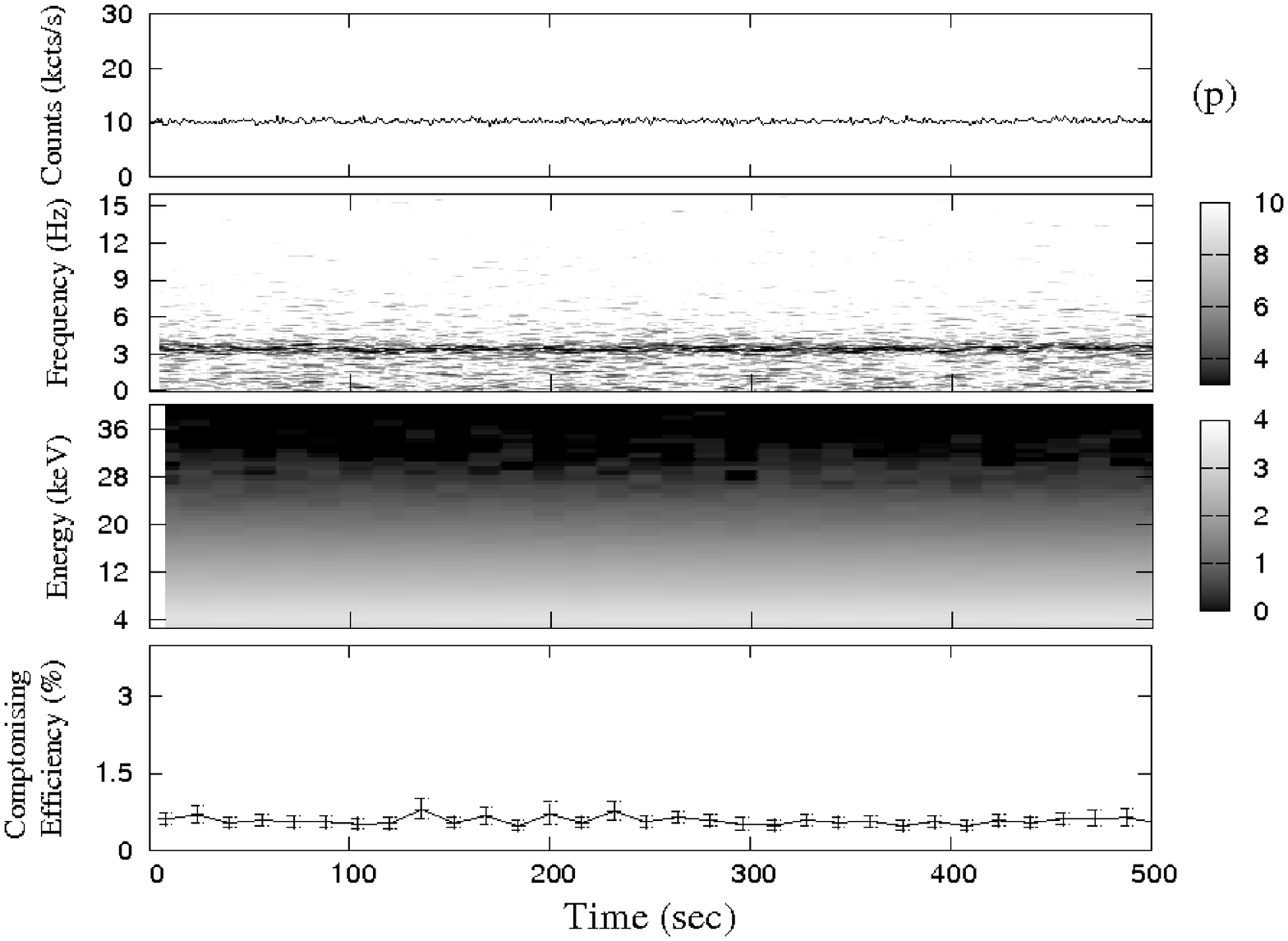}\\
 \includegraphics[angle=270,width=9.50cm]{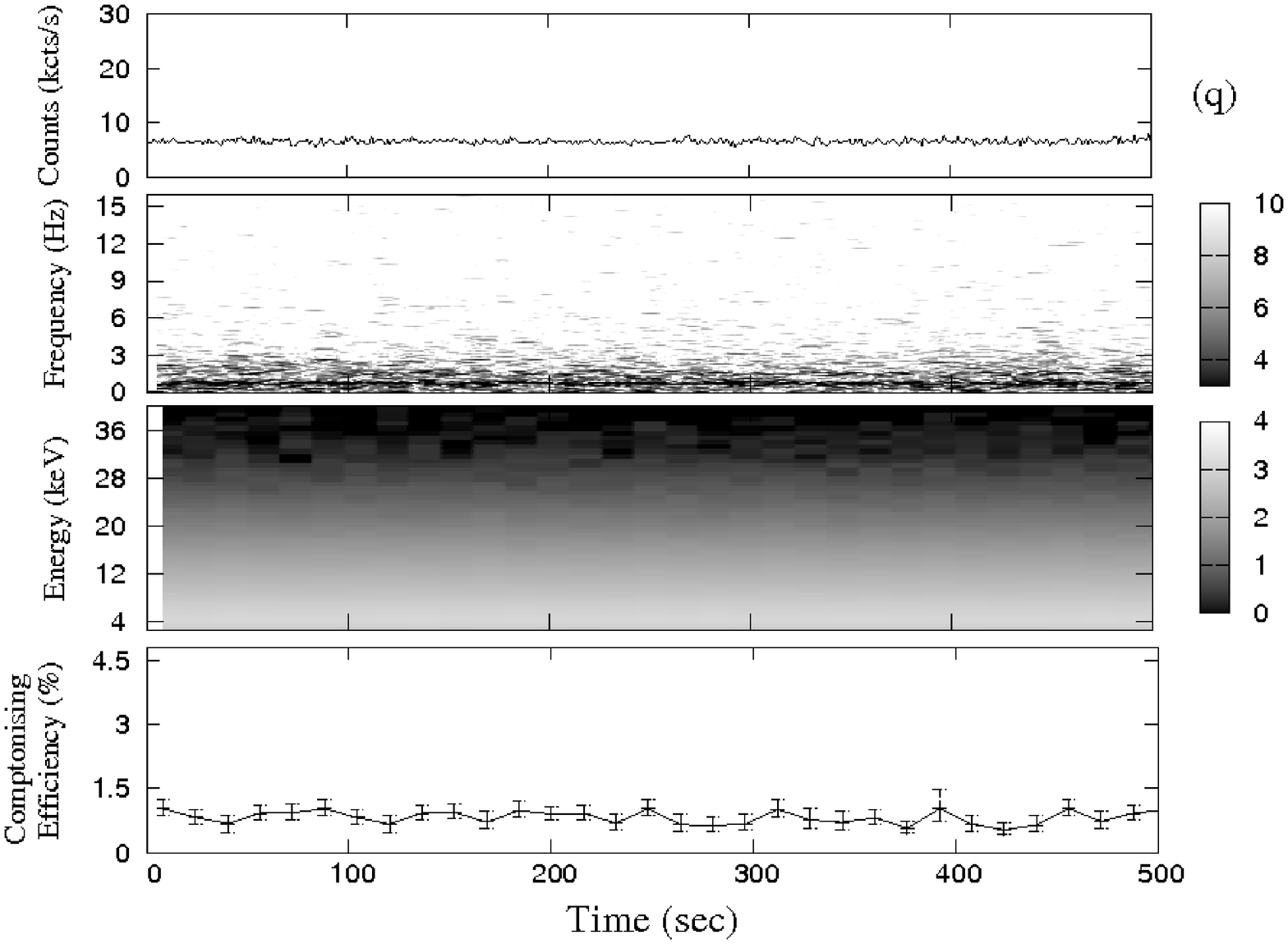}\\
}
\caption{Results of the analysis of (a)  I($\phi$), (b) II($\delta$), (c) III($\gamma$), 
(d) IV ($\omega$),(e) V($\mu$), (f) VI($\nu$),
 (g) VII($\lambda$), (h) VIII($\kappa$), (i) IX($\rho$), (j) X($\xi$), (k) XI($\beta$), (l) XII($\alpha$),
 (m) XIII($\theta$), (n) XIV($\chi_2$), (o) XV($\chi_4$), (p) XVI($\chi_1$), (q) XVII($\chi_3$), 
classes are shown. First Panel: Light curve in $2-40$ keV range, Middle Panel: 
Dynamic power density spectra (PDS) of the light curve. Clear evidence of low frequency noise and 
QPOs are seen. Third panel: Dynamic PCA spectrum showing subtle variations of the 
spectral characteristics with time. Bottom Panel: Comptonizing Efficiency (CE) in 
percent, obtained from $16$s binned data.}
   \label{fig3}
   \end{figure*}

\subsection{Class No. I ($\phi$)}

The Class I data shows a very short time scale variability with the presence of shallow 
dips in its light curve. The results are shown in Fig. 2a.
Dynamic PDS shows no signature of QPOs and the spectrum is mostly soft in nature.
The soft photon rate varies at around $330.24$ kphotons/s while the Comptonized photon rate
varies around $0.16$ kphotons/s. The CE is around $0.05$\%. 

\subsection{Class No. II ($\delta$)}

In Fig. 2b, we show the result of our analysis of the data that belong to the Class II. In this case,
there is an absence of QPO in the dynamic PDS and the spectrum is
soft, mostly dominated by the blackbody photons. The CE is around $0.06$\%. 

\subsection{Class No. III ($\gamma$)}
The class III data appears to be less variable with a distinct and repeated `dip' like features at a 
time gap of a few seconds. The result is shown in Fig. 2c. The class is a steady soft state 
with no QPO visible in the PDS. The CE varies around $0.06$\%. 

\subsection{Class No. IV ($\omega$)}
The results of the analysis of the data of $1000$s in class IV
is shown in Fig. 2d. The photon count rate seems to be steady at around $10000$ 
counts/s for most of the time but sometimes for a duration of 
around $100$ sec the photon count is decreased to about $2000$ counts/sec. 
In the whole class, no QPO is observed and whenever the photon count is 
low, the spectrum appears to become harder. The CE remains low at around $0.07$. 


\subsection{Class No. V ($\mu$)}
The result of our analysis for $500$s of data of class V is shown in Fig. 2e. 
In this class, the spectrum, while remaining soft, shows a considerable 
fluctuation. CE remains low at around $0.05\%$ to $0.15\%$. 

\subsection{Class No. VI ($\nu$)}

The analysis of $500$s of class VI data is shown in Fig. 2f. 
The fluctuations of the photon count rate and the spectrum are erratic.
QPOs are visible. CE varies between $0.04$\% to $\sim 0.3$\%. 

\subsection{Class No. VII ($\lambda$)}

The result of our analysis of $1000$ seconds of data of class VII is shown in Fig. 2g. 
For the first $200-600$ sec the photon number varies between $10000$ and $30000$ counts/s.
The blackbody photon count of the spectrum 
is around $300$ kphotons/s and Comptonized photon rate is around $0.19$ kphotons/s.
This class is a mixture of the burst-off and the burst-on states. When the photon count is higher, 
the spectrum is softer and the object is in the burst-on state
and when the photon count rate is lower, the spectrum is harder and object is in the burst-off state.
The CE varies between $0.05$\% to $\sim 0.1$\%. Here too, the physics of varying the 
Comptonizing region is similar to what is seen in classes VIII-IX below.
At the end of the $400$s span when the burst-off 
state starts, a strong spike in both the CE and QPO frequency are observed,

\subsection{Class No. VIII ($\kappa$)}

An analysis of a $500$s chunk of the data of class VIII of observation 
is shown in Fig. 2h. In the class VIII, the photon counts 
become high $\sim 30000$ counts/sec and low ($\sim 10000$ counts/sec) aperiodically at an 
interval of about $50-75$s.
In the low count regions, the spectrum is harder and the object is in
the burst-off state. Distinct QPOs are present and at the same time, the Comptonizing efficiency 
is intermediate, being neither as high as in the class no. XIII-XVI, nor as low as in class I-III. 
In the fitted spectrum, the Keplerian photon varies between $150$ to $450$ kphotons/sec 
and Comptonized photon varies between $0.39$ to $0.21$ kphotons/sec.
The CE becomes high at $0.25$\%. 

\subsection{Class No. IX ($\rho$)}

The result of our analysis of the class IX data is shown in Fig. 2i. This class contains a 
cyclic variation of the hard and soft photons with roughly $75$s of periodicity. 
The blackbody photon count varies between $350$ to $420$ kphotons/s and Comptonized photons 
vary around $0.95$ to $0.34$ kphotons/s. In the harder states 
CE rises to a maximum of $0.25$\%. 

\subsection{Class No. X ($\xi$)} 

This class analysis is shown in Fig. 2j. In this class, the photon counts are
varying from 2 kcts/s to 8 kcts/s. This is an intermediate class. There
is no distinct QPO visible during this observation. For the first 200 sec
there is a variation in photon counts. But the spectrum remains soft and the CE is 
varying less than 0.1\%. But during 200s-600s the spectrum 
becomes harder and the CE is increased up to 0.3\%. Here we see a correlation in the 
change between the photon counts and the CE variation. During the smaller bumps, the spectrum 
becomes softer and during the bigger bumps, the spectrum becomes harder. 

\subsection{Class No. XI ($\beta$)}

The result of the analysis of $1600$s of Class XI data is shown in Fig. 2k. 
In the first phase of $600$ sec, the 
quasi periodic variation of photon counts takes place with gradually decreasing counts from $10000$
counts/s to $30000$ counts/s. In this phase, the spectrum is soft and the QPO is seen only when 
the photon count is low. The CE varies around $0.14\%$. 

In the next $800$ sec the spectrum is harder and the distinct variation of QPO frequency (from 
$12$ Hz to $3$ Hz) indicates the variation of the shock location. However, in this phase,
CE is around $0.27\%$. 

\subsection{Class No. XII ($\alpha$)}

Class XII is an intermediate class in which QPO is always observed, 
the QPO frequency is generally correlated with the count rate (as in
XIV-XVII classes discussed below). Since this class displays a long time variation,
we analyze the data of $2000$s.  Fig. 2l shows the result. 
The CE varies between $0.05$\% to $\sim 0.6$\%.

\subsection{Class No. XIII ($\theta$)}

The result of the analysis of a $1000$s data of class XIII is shown in Fig. 2m. This
class can be divided in two regions depending on the photon count rates. In the soft dip region the 
photon count rate is lower than $10000$ counts/s. In this region, the spectrum is softer
and the CE is around $0.17\%$. 

In the other (hard dip) region, say, between $350$s and $820$s, the photon count is higher and 
varies from $10000$ counts/s to $30000$ counts/s. In this region, CE reaches a high value of $0.28$\%. 

\subsection{Classes No. XIV ($\chi_2$) \& XV ($\chi_4$)}

XIV and XV classes have low soft X-ray fluxes with less intense radio emissions. 
Dynamic PDS shows a distinct QPO at around $3$ Hz in class XIV and around $5$ Hz in class XV. 
The X-ray photon is significant up to $40$ keV with a flatter power-law slope, which signifies that the 
source belongs to a harder state. The results are given in Fig. 2n and Fig. 2o.
The Comptonizing efficiency in XIV and XV classes is around $0.4-0.5$\%. 

\subsection{Classes No. XVI ($\chi_1$) \& XVII ($\chi_3$)}

The classes XVI and XVII correspond to the radio-loud
states, whereas the classes XIV and XV are in radio-quiet states \citep{vad01}. 
In both the cases, the dynamic PDSs show a strong QPO feature around 
$4$ Hz and $0.8$ Hz respectively throughout the particular observation and the spectrum is dominated 
by the hard power-law. Results are shown in Fig. 2p and Fig. 2q. Around $0.77$\% of 
the soft photons are intercepted by the CENBOL in XVI class, whereas it is $0.88$\% in XVII class.

\section{Comptonizing efficiency in different classes: an unifying view}

In the above Sections we have analyzed the data of all the seventeen classes. 
We presented the light curves, the dynamical energy spectra, the power density spectra,
and the ratio of the power-law photon rates to the blackbody photon rates obtained from the fit
of the original spectra. 

To interpret the results, and to facilitate the discussions, it is instructive 
to keep a paradigm in mind. In Figs. 3(a-c), we present a cartoon diagram which is basically
the two component advective flow (TCAF) model of \citet{skc95},
suitably modified to include outflows as in \citet{skc00b}. 
In the Figure, the BH represents the black hole, the dark shaded disk is the Keplerian 
flow and the light shaded disk is the sub-Keplerian flow. The CENBOL, the centrifugal pressure supported
boundary Layer of the black hole, and the outflow emanated from it intercept some
soft photons from the Keplerian disk and may be cooled down if the Keplerian rate and/or
the degree of interception is sufficient. 

\begin{figure*}
\centering
\includegraphics[angle=0,width=13.0cm]{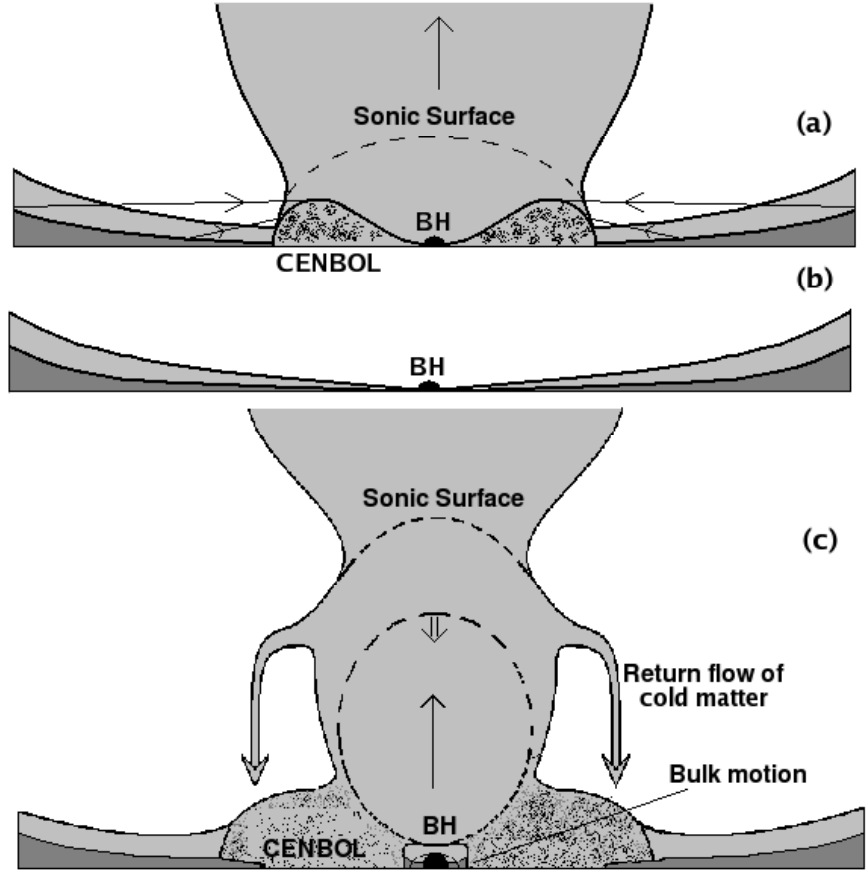}
\caption{Cartoon diagrams for three major types of variabilities in GRS 1915+105.
The classes XIV-XVII belong to the group (a) where the Keplerian rate is low, the CENBOL is large
and $<CE>$ is high. Jet/outflow rates may be low and continuous. This is the so-called `Hard' group. 
The softer classes such as I-IV belong to group (b) where the CENBOL is very small and very little 
Comptonization which could be taking place is 
due to the sub-Keplerian flows above the Keplerian disk. This is the `Soft' group. 
The rest of the classes belong to the variation of group (c) or the `Intermediate' group, 
where, the jet also plays dynamically important role in shaping the spectrum \citep{skc00b}.
\label{proof}}
\end{figure*}

In the cartoon diagram of Fig. 3a, the CENBOL is not cooled enough and the Comptonization
of the soft photon is done both by the CENBOL as well as by the outflow. This configuration typically 
produces a `harder' state. However, the outflow rate depends on the shock strength \citep{skc99}
and could produce weak jets as in XIV-XV or strong jets as in 
XVI and XVII \citep{vad01}. In Fig. 3b, the Comptonization is
due to the high accretion rate in the Keplerian disk and the CENBOL collapses. 
As a result, no significant jet or outflows is expected. This configuration typically produces 
the softer states. If both the Keplerian and the sub-Keplerian rates 
are comparable, then the intermediate states would be produced also. 
In Fig. 3c, we show the situation where the shock strength is intermediate and consequently, 
the outflow rate is the highest (see, \citet{skc99, skc00b} and references therein). 
In this case, there is a possibility that the outflow may be cooled down 
by Comptonization if the intercepted soft photon flux is high enough and 
the outflow is temporarily terminated. The flow falls back to 
the accretion disk, increasing the local accretion rate in a very short 
time scale (seconds) \citep{skc00a, skc00b}.
The possibility of a similar change in local accretion rate 
was mentioned also by \citet{cab09}.  We believe that the variability classes 
(such as V-XIII) some of which show clear softer (burst-on) 
and harder (burst-off) states alternately and showed evidences of intermittent 
jets \citep{wolt02, rod08} belong to this category \citep{skc00a}.

In TCAF paradigm, results obtained from the analysis are also interpreted. 
The result obtained for Class No. I($\phi$) and Class No. II($\delta$) means that a very few 
black body photons are intercepted by the CENBOL. This corresponds to a situation 
similar to that in Fig. 3b.  
In case of Class No. III($\gamma$) the analysis results indicate a low degree of interception 
of the soft photons with the CENBOL.
The four classes I-IV belong to the softer class. The CE is very low, even when the
Keplerian rate is high (spectrum is dominated by soft photons). This means that the CENBOL is 
very small in size and is steady (cooling time scale is much shorter than the infall time scale)
and hence the QPOs are also absent. In TCAF paradigm, these cases would correspond to
that in Fig. 3b, only the accretion rates vary. From the count rates, it appears that the disk rates are 
intermediate in classes I and II while it is high in class III. The sub-Keplerian rate is low in 
class II, but is intermediate in classes I and III.
In case of Class No. V ($\mu$) the Keplerian rate may be changing rapidly 
and the shock is not really formed. This happens when the 
sub-Keplerian flow has a very low energy and/or
angular momentum \citep{skc90}.
In Class No. VII ($\lambda$) we can observe an indication of 
sudden change in the Compton cloud optical depth.
In Case of Class No. VIII ($\kappa$) burst-off state, low frequency QPOs are indicators of 
the shock oscillations \citep{skc00a}. As discussed before, in this case, the 
strength of the shock is intermediate and produces strong outflows \citep{skc00a}. 
In case of burst-on state of the same class the CENBOL would be cooled down. 
Some matter returns back from the wind on the CENBOL. When it is totally drained, 
the burst-off state is resurfaced with a lower CE (less than $0.1\%$). 
In case of Class No. XI ($\rho$) the harder states
(low count regions), the CENBOL is prominent 
and the Keplerian component is farther away. Thus the QPO is prominent.  
Interception of black body photons increases with CENBOL size and 
CE rises to a maximum of $0.25$\%. 
As the Keplerian disk moves towards the black hole, a gradual softening 
of the spectrum occurs since the CENBOL becomes smaller in size while still retaining its identity. 
At the peaks of the light curve, the CENBOL, which is also the base of the jet is cooled down
due to the increased optical depth.
In case of Class No. XI ($\beta$) the outflow is taking an active
role in intercepting the soft-photons. We suspect that sudden rise in magnetic activity 
occurs in this case which collapses the CENBOL due to increased magnetic tension in the hotter
plasma \citep{nan01b}.
In case of Class No. XII ($\alpha$) the spectrum shows that the Keplerian rate 
is not changing much, but the sub-Keplerian flow
fluctuates, perhaps due to failed attempt to produce sporadic jets. This case may also
require parameters outside the scope of TCAF to understand fully.
For Class No. XIII ($\theta$) 
The QPO is also present. Here, the CE, which is linked to the 
geometry of the Comptonizing region, is anti-correlated with the QPO 
frequency. In TCAF paradigm, QPO this is correlated with shock location, i.e., the CENBOL size. 
In the third panel, where the dynamic spectrum is drawn, we note that between the soft dip and
the hard dip, the low energy photon intensity is not changing much, while the intensity of 
spectrum at higher energies is higher. This indicates that the sub-Keplerian rate is high
in the hard dip state. Here too the presence of a strong magnetic field may be required 
to understand this class fully.
In Class No. XIV ($\chi_2$) and XV ($\chi_4$) he CENBOL is big in size and the 
Keplerian disk is farther out and/or with a very low rate.
In case of Class No. XVI ($\chi_1$) and XVII ($\chi_3$) according to TCAF paradigm, 
the CENBOL (Fig. 3a) oscillates to produce the observed QPOs \citep{skc97, skc00a}. 
The Keplerian rate is low and is unable to cool the CENBOL and the jet combined.
Whether the outflow rate would be high or low would depend on the shock strength \citep{skc99}.

In the literature, to our knowledge, there has been no discussion on the sequence 
in which the class transitions should takes place. Also, there is no discussion on
which physical properties of the flow the variability classes depend. 
In the above, we analyzed several aspects of the variability classes 
which may be understood physically from the two component advective disk paradigm.
Since CE behaves differently in various classes, it may play
an important role in distinguishing various classes. This CE
is directly related to the optical depth of the Compton cloud, since the optical depth 
determines the fraction of the injected soft photon intercepted by the Comptonizing region.
   \begin{figure*}
   \centering
   \includegraphics[angle=-90,width=7cm]{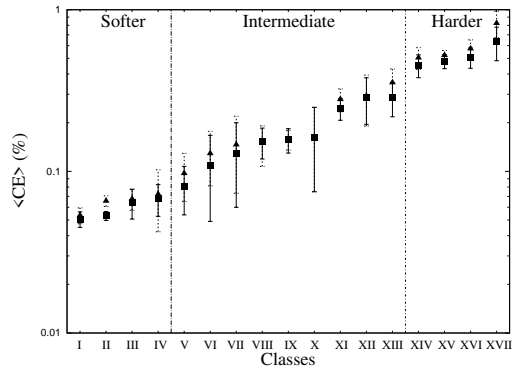}
\caption{Variation of the average Comptonizing efficiency ($<CE>$) for different variability classes 
of GRS 1915+105. The error-bars were calculated from the excursion of CE in a given class. Filled squares and 
triangles represent $<CE>$ for two sets of variability classes. Softer classes have smaller $<CE>$ and
harder classes have higher $<CE>$.
\label{seq}}
     \end{figure*}

Though CE varies in a given class, it is instructive to compute the average  
CE in a given class (averaged over a period characterizing the class). The average is computed as follows:
$\langle CE \rangle = \Sigma(CE_h Dt_h +CE_s Dt_s)/ \Sigma(Dt_h+Dt_s)$, where, the subscripts $h$ and $s$ under CE denote 
its value when the spectrum is harder (for a residence time of $Dt_h$) and softer (for a residence 
time of $Dt_s$) respectively. $Dt_h+Dt_s$ denotes the full cycle time of a given class. We denote this as $<CE>$. 
In Fig. 4, we present the variation of log($<CE>$) in various classes. We placed `error bars'
also in all the average values which are at 90\% confidence level. These `error-bars'
are actually the excursion of $<CE>$ in that class, and not actually the error-bars in the 
usual sense. We arranged the classes in ascending order of $<CE>$.
This gives rise to a sequence of the variability class shown in the 
X-axis. In classes XI and XIII, we did not average CE over both hard and soft  
regions, since it is believed that the soft regions are produced due to sudden
disappearance of the Compton cloud \citep{nan01a}.   
Thus, while placing them in the plot, we considered the average over the 
burst-off (harder) state only. To show that the sequence drawn is unique, we plotted the  
average values of CE of two sets of data (Table 1). The average of 
the first set is drawn with a dark square sign and the average of the 
second set is drawn with a dark triangle sign. The individual error bars are 
drawn with solid and dashed lines respectively. In both the sets, the sequence is identical.
Based on the nature of the variations of CE, we divided these classes into three groups corresponding to three
types of accretion shown in Figs. 3(a-c). Classes I to IV belong to the `softer' states
and classes XIV-XVII belong to the `harder' states. The rest of the classes belong to the `intermediate' states.

Once we sequence the classes based on $<CE>$, one could ask if the sequence means anything, physically and 
observationally. Physically, increasing average $<CE>$ corresponds of increasing optical depth of the 
Comptonizing region and its excursion signifies how the optical depth varies in short timescales, i.e.,
accretion rate in the low-angular momentum component (Fig. 3). Observationally, only a handful
class transitions have been caught `red-handed'. About them we shall discuss later below.
   \begin{figure*}
   \centering
   \includegraphics[angle=-90,width=6.5cm]{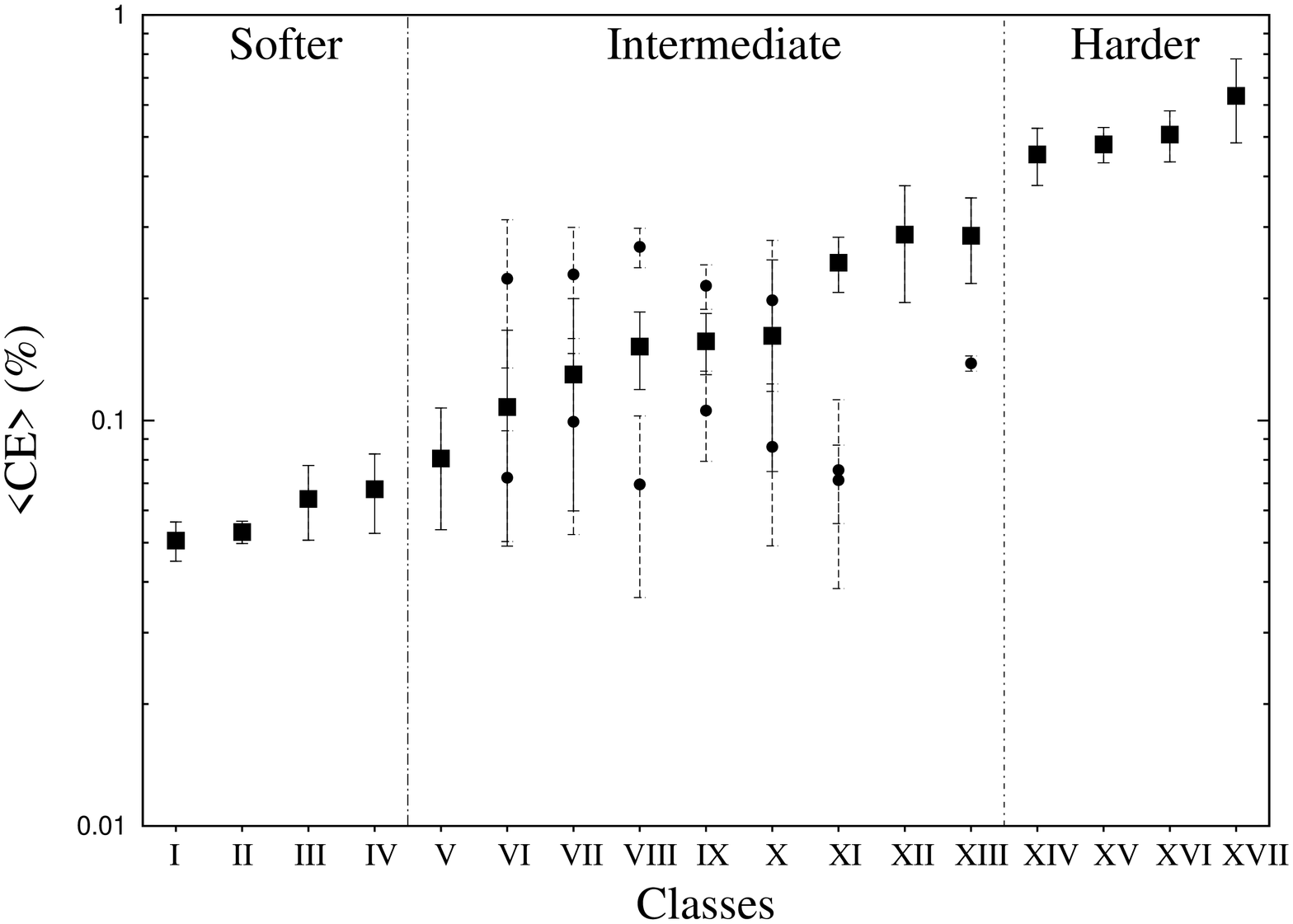}
   \includegraphics[angle=-90,width=6.5cm]{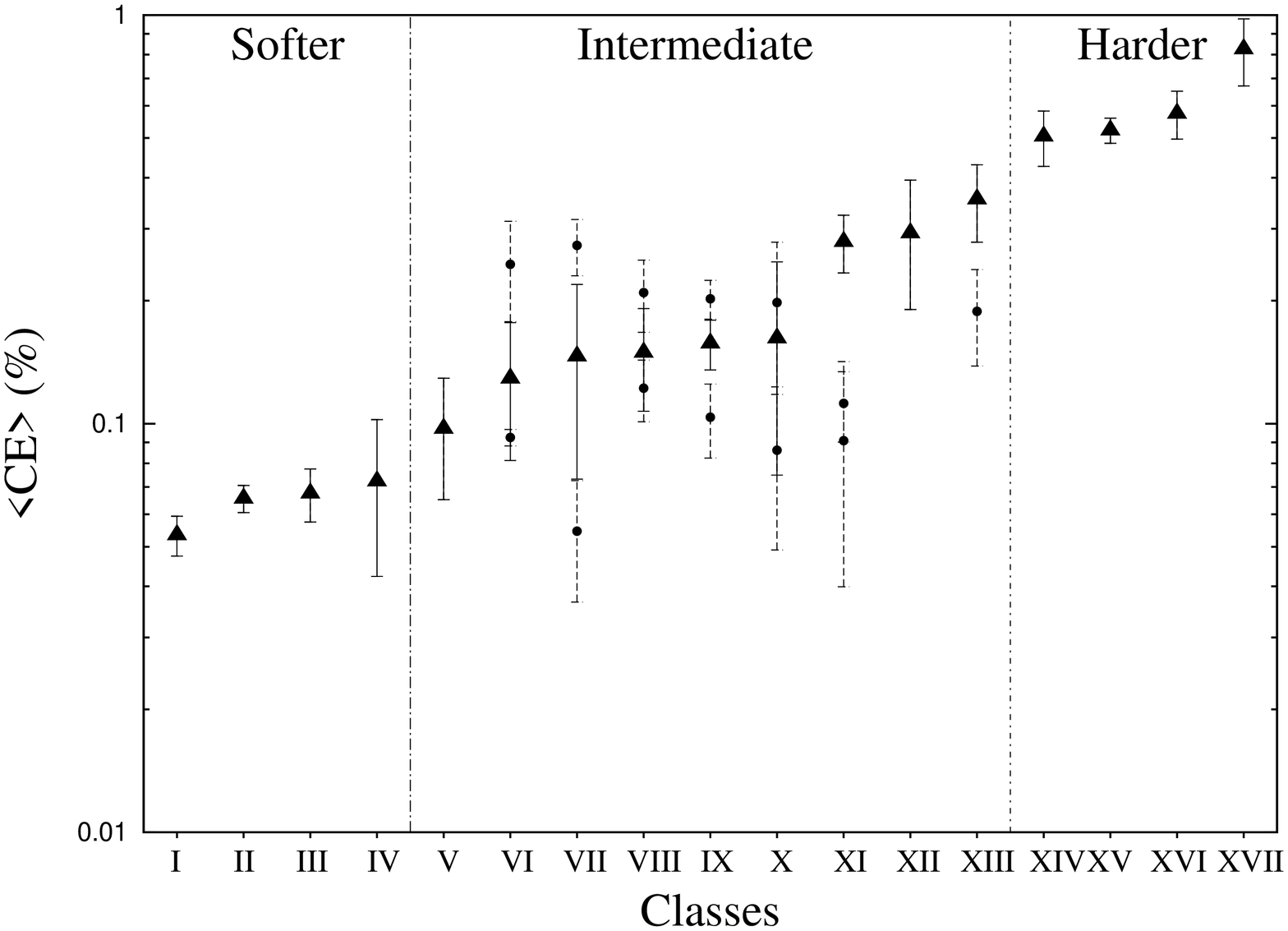}
\caption{Variation of average Comptonizing efficiency ($<CE>$) for different variability 
classes of GRS 1915+105.
In the left panel, we show the results for the data set plotted with filled squares in Fig. 4,
and the right panel, we show the results of the
data plotted with filled triangles in Fig. 4. The averages in hard
and soft chunks are placed separately with filled circles. In class XI
averages of three chunks (burst on, pre-spike and post-spike) are plotted with filled
circle, filled box and filled circle respectively.
The overall sequence is found to remain the same in both sets of data. We placed classes XI and XIII
according to $<CE>$ in the hard (pre-spike) region since the softer regions
are believed to be produced due to totally different physical processes.}
\label{seq2}
     \end{figure*}

In Figs. 5(a-b), we plotted the results of these two sets of analysis separately. 
We separate the averages over CE in the burst-off and burst-on states 
(as dark filled circles) when they are present as well as the global average 
over CE in a given class by filled squares (except classes XI and XIII where the physics 
is different and it is meaningless to talk about the overall average; and hence only 
average over the harder region is plotted.).

In \citet{skc04, skc05}, it was mentioned 
that although many RXTE observations were made of GRS 1915+105, only in a few cases, 
direct transitions were observed. In particular, using Indian X-ray Astronomy 
Experiment (IXAE) it was shown that the direct transitions of $\kappa \rightarrow \rho$ (VIII $\rightarrow$ IX), 
$\chi \rightarrow \rho$ (XIV $\rightarrow$ IX), $\chi \rightarrow \theta$ (XIV$\rightarrow$ XIII) and 
$\rho \rightarrow \alpha$ (IX$\rightarrow$ XII) do take place in a matter of hours. In \citet{nai02b}, IXAE
data was used to argue that $\rho$ class variability could have changed 
to $\chi$ class via $\alpha$ class (i.e., IX $\rightarrow$ XII $\rightarrow$ XIV). In \citet{nan01b}, it was 
shown that the $\theta$ class (Class XIII) is anomalous  and the observed soft-dip is perhaps due to the disappearance 
of the CENBOL by magnetic rubber band effect (this is the magnetized TCAF or 
MTCAF model). Accepting the class XIII to be anomalous, we find from Fig. 4 that the observed 
transitions reported in \citet{skc04, skc05, nai02b} are `naturally' explained. 
For instance, $\chi \rightarrow \rho$ ($\theta$ and $\beta$ being anomalous and intermediate
$\alpha$ has been reported by \citet{nai02b}), $\rho \rightarrow \alpha$
($\beta$ being anomalous) and $\kappa \rightarrow \rho$, 
are expected from our analysis. We believe that if we carry out spectro-photometry 
of GRS 1915+105 continuously \citep{skc08}, then
we may be able to catch the transition from one type to another more often and check if the 
sequences we mentioned here are sufficiently robust or need further refinement.

   \begin{figure*}
   \centering
   \includegraphics[angle=-90,width=7cm]{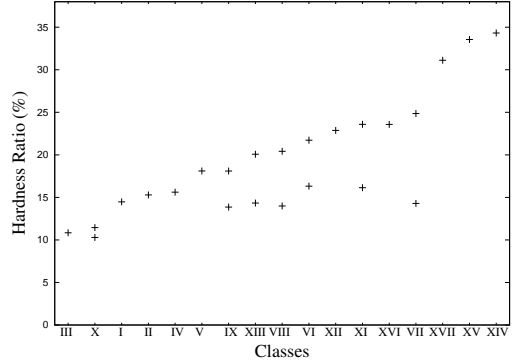}
\caption{Variation of the hardness ratio for different variability classes of GRS 1915+105. 
The hardness ratio is the ratio between (10 - 40 keV) and (2 - 10 keV) photon counts in $\%$.
A sequence of variability classes in X-axis is obtained by placing the hardness ratios in ascending order. 
For intermediate states the hardness ratio of the harder portions is used for the sequencing of 
the classes. This sequence is totally different from that presented in Fig. 5 and is not connected to
any observed transition.
\label{hr}}
     \end{figure*}

   \begin{figure*}
   \centering
   \includegraphics[angle=-90,width=7cm]{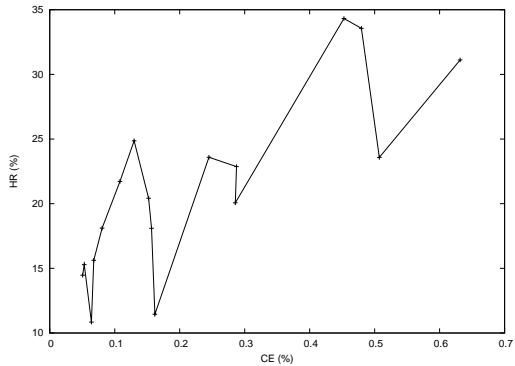}
\caption{Variation of HR($\%$) vs CE($\%$) of GRS 1915+105. The definition of HR is same as given in Fig. 6. 
CE is plotted is ascending order but the variation of HR do not follow the same sequence along with the variation of CE.  
\label{hvsce}}
     \end{figure*}

In Fig. 6, we plotted the variation of the conventional hardness ratio for all the classes of GRS 1915+105 for
the sake of comparison of CE. Here, the hardness ratio is the ratio (in $\%$) between the photon rates in (10 - 40 keV)
and in (2 - 10 keV) range. The classes are arranged in the ascending order of hardness ratios. 
In case of intermediate classes we have taken the hardness ratio of the harder portions.
This sequence obtained from this Figure does not seem to have any special significance in that
it does not support the observed class transitions reported in \citet{skc04, skc05}.  

In Fig. 7, we plotted the variation of the variation of conventional HR for all classes along with the CE for GRS 1915+105.
This Figure shows that the variation of HR do not follow one to one correspondence with the variation of CE.

The conventional hardness ratio is generic as it is not based on any radiative processes. It is independent of 
even the black hole mass and accretion rates. That is the main reason why the CE defined by 
us may be considered to be more physical.

\section{Conclusions}

In this paper, we have analyzed all known types of variability classes of the enigmatic black hole GRS1915+105
and computed the dynamical nature of the energy and the power density spectra. We did not characterize
these classes by conventional means, such as the hardness ratios defined in certain energy range since
such a characterization does not provide us with any insight about the physical picture. 
Furthermore, characterization using conventional hardness ratio can be done on a case by case basis, 
and cannot be valid for the black holes of all masses. 
Because of this we concentrated on the ratio of the photon fluxes in the 
power-law and black body components as seen in the disk frame. 
This is a quantity similar to the conventional hardness ratio,
but the energy ranges automatically vary from class to class as obtained by minimizing $\chi^2$. 
We asked ourselves whether we can distinguish one class from another 
from physical point of view independent of what the nature of the Compton cloud is. 
We computed the mean $<CE>$ Comptonizing efficiency and put them in an ascending order.
This sequence remains the same in two independent sets of 
data we studied. So $<CE>$ may be thought of characterizing a class. Based on the values of $<CE>$, 
it is observed that the classes belong to three states: Classes I-IV belong to the softer states, 
Classes V-XIII belong to the intermediate states, and Classes XIV-XVII belong to the harder state.

Surprisingly, the handful of class transitions in GRS 1915+105 which have been caught `red-handed' 
appear to take place between the nearest neighbors of this sequence. It is 
hoped that the transitions observed in future may also take place 
between two nearest neighboring classes unless the neighboring 
class itself is anomalous (such as, $\theta$, $\beta$ and $\alpha$  
which appear to require enhanced magnetic activities). 
If so, then it would imply that the average optical depth 
of the Compton cloud (which is decided by the specific angular momentum and the accretion rate of 
the low-angular momentum halo, according to \citet{skc95} model) holds an important 
key to decide a variability class. Neither the accretion rate of the 
Keplerian disk nor the accretion rate of the low-angular
momentum matter alone individually can be considered to be the deciding factor. 

In the two component model of \citet{skc95}, the CENBOL and the associated outflow 
play the role of the Compton cloud. In this model, a change in CE
can be achieved in several ways: (i) A change in the shock location (achieved by changing
primarily the specific angular momentum) change the size of the CENBOL and thus the 
degree of interception of soft photons, and/or,
(ii) by changing the accretion rate of the sub-Keplerian component, which changes
the optical depth and scatter a different number of soft photons to produce the power-law photons. 
However, the survival of the Compton cloud depends on the accretion rate in the Keplerian 
component. If the rate is higher, the Compton cloud collapses. Thus, CE is the result of
non-linear relationships among the flow parameters. In \citet{skc95, skc97} the spectral state variation 
was shown to be due to these effects. In our present situation, we find that the classes with 
harder states have more CE. However, since shifting the shock location or raising the
sub-Keplerian rate does not involve viscous effects, changes in softer to harder states or vice versa
can be achieved in minutes or even seconds. The time scales of such effects through a Keplerian flow
could take days. Thus, we believe that not only we identify the most important parameter
which distinguishes from one class to the other, we also believe that it is understandable within 
the framework of the TCAF model.

While we obtain a hint that the $<CE>$ is important in deciding the class transition sequence,
for a given class, the degree of excursion of CE in a 
given variability type must depend on another parameter, such as the outflow rate. 
In the TCAF picture, a simple estimate \citep{skc99} suggests that the shock 
strength decides the outflow rate and therefore the time taken by the 
base of the outflow to reach $\tau \sim 1$ for a Compton scattering to take place \citep{skc00a}.
Thus the outflow rate also decides the time-scale for which a class will stay in a harder or
softer states, which in turn, decides the degree of excursion of CE in a given class.
For a very strong shock, the outflow rate is very weak, as is evidenced by a weak 
radio flux even for `radio-loud' classes. For intermediate shock strength 
the outflow rate is higher, and it is easier to have $\tau=1$ in a short time scale \citep{skc00a}.
The fractional change in CE can also become high. In classes XIII-XVII, we not only find
CE to be very high, the excursion of CE is very small as well.
The aspect of classification in terms of the outflow rate is being looked into. 
The analysis is in progress and will be reported elsewhere.

It is to be noted that GRS 1915+105 has been observed, though intermittently, for over 17 years now and 
we have analyzed only few tens of data sets, each lasting for $\sim 2000$s. Thus it is likely that more 
exhaustive data sets or future observations may force us to re-look at the sequence that we prescribe here.
One limitation of our work is that we do not include the effects of the hardening factor of the spectra 
in separating the power-law and the black body photons. With the present understanding of the anomalous 
classes such as $\beta$ and $\theta$, it is not yet clear why sometimes
these classes are skipped while transiting from one class to another, other than to 
speculate that these classes require a new parameter, such as strong magnetic fields, which may 
be due to magnetic activity of the companion star.
With an improvement in understanding a further refinement of our sequence may not be ruled out.

\section{Acknowledgment}
PSP acknowledges SNBNCBS-PDRA Fellowship.


\end{document}